\documentclass[3p,times,authoryear]{elsarticle}

\usepackage{soul}

\usepackage{amssymb}
\usepackage{booktabs}
\usepackage{color}
\usepackage{makecell}
\usepackage{tikz}
\usepackage{amsmath}
\usetikzlibrary{arrows.meta, positioning, calc}
\usetikzlibrary{shapes.geometric}
\usepackage{tcolorbox}
\usepackage[table]{xcolor}
\usepackage{subcaption}
\usepackage{calc}
\usetikzlibrary{calc}
\usetikzlibrary{shapes,arrows,positioning,calc,decorations.pathreplacing}

\usepackage{longtable}
\usepackage{array}
\usepackage{xcolor}
\usepackage{multirow}
\usepackage{geometry}
\definecolor{KA1}{RGB}{140,86,75}  
\definecolor{KA2}{RGB}{44,160,44}   
\definecolor{KA3}{RGB}{214,39,40}   
\definecolor{KA4}{RGB}{148,103,189} 
\definecolor{KA5}{RGB}{31,119,180}  
\definecolor{KA6}{RGB}{188,189,34} 
\definecolor{KA7}{RGB}{127,127,127} 
\definecolor{KA8}{RGB}{227,119,194} 
\definecolor{KA9}{RGB}{255,127,14}  
\newcommand{\graybg}{\rowcolor{gray!15}}

\usepackage{pgf-pie}  
\usetikzlibrary{fit}

\newcommand{\cmark}{\ding{51}}%
\newcommand{\xmark}{\ding{55}}%
\usepackage[normalem]{ulem}

\newcommand{\CLLM}{CurricuLLM}

\newcommand{\piechart}[3][0.5]{
    \begin{tikzpicture}[baseline=-0.5ex]
        \def\radius{#1}
        \pgfmathsetmacro{\total}{0}
        
        \foreach \value/\color in {#2} {
            \pgfmathparse{\total + \value}
            \global\let\total\pgfmathresult
        }
        
        \pgfmathsetmacro{\startangle}{0}
        
        \foreach \value/\color in {#2} {
            \pgfmathsetmacro{\endangle}{\startangle + \value/\total*360}
            \draw[fill=\color, draw=black] 
                (0,0) -- (\startangle:\radius) 
                arc (\startangle:\endangle:\radius) -- cycle;
            \global\let\startangle\endangle
        }
        
    \end{tikzpicture}
}

\newcommand{\secref}[1]{Section~\ref{sec:#1}}

\newcommand{\figref}[1]{Figure \ref{fig:#1}}

\newcommand{\tabref}[1]{Table~\ref{tab:#1}}

\usepackage[figuresright]{rotating}
\usepackage{url}
\usepackage[hidelinks]{hyperref}

\begin{document}

\begin{frontmatter}





\title{\CLLM: Designing Personalized and Workforce-Aligned Cybersecurity Curricula Using Fine-Tuned LLMs}

\author[label1]{Arthur Nijdam (corresponding author)}
\ead{arthur.nijdam@eit.lth.se}
\author[label2,label3]{Harri Kähkönen}
\ead{harri.kahkonen@helsinki.fi}
\author[label2,label3]{Valtteri Niemi}
\ead{valtteri.niemi@helsinki.fi}
\author[label1]{Paul Stankovski Wagner}
\ead{paul.stankovski\_wagner@eit.lth.se}
\author[label1,label4]{Sara Ramezanian}
\ead{sara.ramezanian@kau.se}

\address[label1]{Lund University, Department of Electrical and Information Technology, Box 118, SE-221 00 Lund, Sweden}
\address[label2]{University of Helsinki, Department of Computer Science, Helsinki, Finland}
\address[label3]{Helsinki Institute for Information Technology, Helsinki, Finland}
\address[label4]{Karlstad University, Department of Mathematics \& Computer Science, Karlstad, Sweden}







\begin{abstract}


The cybersecurity landscape is constantly evolving, driven by increased digitalization and new cybersecurity threats. Cybersecurity programs often fail to equip graduates with skills demanded by the workforce, particularly concerning recent developments in cybersecurity, as curriculum design is costly and labor-intensive.  
To address this misalignment, we present a novel Large Language Model (LLM)-based framework for automated design and analysis of cybersecurity curricula, called \CLLM{}. Our approach provides three key contributions: (1) automation of personalized curriculum design, (2) a data-driven pipeline aligned with industry demands, and (3) a comprehensive methodology for leveraging fine-tuned LLMs in curriculum development. 

\CLLM{} utilizes a two-tier approach consisting of PreprocessLM, which standardizes input data, and ClassifyLM, which assigns course content to nine Knowledge Areas in cybersecurity.
We systematically evaluated multiple Natural Language Processing (NLP) architectures and fine-tuning strategies, ultimately selecting the Bidirectional Encoder Representations from Transformers (BERT) model as ClassifyLM, fine-tuned on foundational cybersecurity concepts and workforce competencies.

We are the first to validate our method with human experts who analyzed real-world cybersecurity curricula and frameworks, motivating that \CLLM{} is an efficient solution to replace labor-intensive curriculum analysis.  Moreover, once course content has been classified, it can be integrated with established cybersecurity role-based weights, enabling alignment of the educational program with specific job roles, workforce categories, or general market needs. 
This lays the foundation for personalized, workforce-aligned cybersecurity curricula that prepare students for the evolving demands in cybersecurity. 


\end{abstract}

\begin{keyword}
Cybersecurity, Curriculum Design, Large Language Models, Personalized Learning, Education 


\end{keyword}

\end{frontmatter}



\section{Introduction} \label{sec:introduction}

Society has become more dependent on digital systems due to, e.g., smart cities \citep{gibson1992technopolis}, Internet of Things \citep{rath2024role} and Artificial Intelligence (AI) \citep{turing2004intelligent}. This has resulted in a corresponding surge in sensitive data, an increasing demand for the development and deployment of digital devices, and a persistent need for user education and training. A fundamental pillar in sustaining the digital ecosystem is cybersecurity. Cybersecurity professionals act in protecting against cyber-attacks \citep{lehto2022cyber}, assessing cyber risks \citep{KHALIL2025104015}, safeguarding user privacy \citep{barth2022lost}, protecting minors in online environments \citep{ramezanian2019privacy}, and fostering user trust \citep{barth2022lost}. Despite its critical importance, the estimated cybersecurity workforce gap was at 4.8 million globally in 2024 \citep{ISC2}.  


At the same time, educational programs struggle to keep up with the fast-paced changes in the cybersecurity landscape \citep{blavzivc2022changing, aldaajeh2022role, catal2023analysis, ramezanian2024cybersecurity}. This dynamic landscape, coupled with the inherently multidisciplinary nature of the field, underscores a growing demand for personalized learning paths \citep{cabaj2018cybersecurity}. Personalized curricula enable students to either specialize in their area of interest or gain broad knowledge across multiple disciplines. 
Nevertheless, several challenges persist: 

\begin{itemize}
    \item There are large differences between institutions regarding which components of cybersecurity education should be classified as `core' or `elective' \citep{cabaj2018cybersecurity}. Core curricula might lack essential topics relevant to industry. 
    \item Increasing responsibility is placed on the student to select `the right' courses for their future career. However, research shows that students primarily tend to choose electives based on their perceived difficulty, instead of the student's envisioned career \citep{ting2012understanding}. 
    \item Adequate guidance counseling should be available to advise students on their curriculum \citep{noaman2015new}, which requires significant resources.  
    \item Ensuring that curricula remain aligned with recent developments and industry demands presents a significant challenge for educators and curriculum developers \citep{blavzivc2022changing}. 
\end{itemize}

To address these challenges, we propose \CLLM{}, a novel end-to-end framework for cybersecurity curriculum analysis based on fine-tuned Large Language Models (LLMs).\footnote{It is worth noting \cite{robotCurricuLLM} also proposed a system named \CLLM{}. Their work revolves around curriculum learning, a technique used within Reinforcement Learning where agents are trained on progressively harder environments, and is unrelated to our approach.} LLMs excel at processing text, making them particularly suitable for analyzing and interpreting educational guidelines and course material. Prior work has demonstrated the proficiency of these models in processing disciplinary cybersecurity knowledge \citep{zhang2025llms, tihanyi2024cybermetric}.

By harnessing the language comprehension capabilities of LLMs, \CLLM{} automates the assessment of cybersecurity curricula against standardized educational and workforce frameworks. While our current implementation utilizes the Cybersecurity Curricula 2017 (CSEC2017) \citep{csec2017} and the 2025 National Initiative for Cybersecurity Education (NICE) workforce framework \citep{NICE2025} as illustrative examples, \CLLM{} is designed to flexibly support other competency or knowledge frameworks. That is, our \CLLM{} pipeline is built to be easily adapted to other workforce or curriculum frameworks with similar structure, even if the domain is unrelated to cybersecurity.

\CLLM{} enables mapping knowledge requirements of cybersecurity job roles (defined in the NICE framework) to corresponding Knowledge Areas (defined in the CSEC2017 framework), assisting educators, guidance counselors and students in identifying knowledge gaps, aligning learning outcomes with evolving skill demands, and tailoring the students' skill set towards a desired career path. Ultimately, \CLLM{} contributes to more effective and targeted cybersecurity education.

\begin{table}[h]
\centering
\caption{Summary of notations, in alphabetical order.}
\begin{tabular}{p{3cm}p{10cm}}
\hline
\textbf{Abbreviation} & \textbf{Description} \\
\hline
AI & Artificial Intelligence  \\
BERT &  Bidirectional Encoder Representations from Transformers \citep{devlin2019bert} \\
\CLLM{} & automated curriculum assessment tool consisting of PreprocessLM and ClassifyLM \\ 
ClassifyLM & Language Model used to classify data into 9 Knowledge Areas \\
CSEC2017 & Cybersecurity Curricula 2017 framework \citep{csec2017} \\
DCWF & DoD Cyber Workforce Framework \citep{doddcwf} \\
ENISA & European Union Agency for Cybersecurity  \\
KA  & Knowledge Area, defined in the CSEC2017 framework \\
KD   & Knowledge Description, defined in the NICE framework\\
KU  & Knowledge Unit, defined in the NICE framework  \\
LLM & Large Language Model \\
NICE & National Initiative for Cybersecurity Education \citep{NICE2025}, developed by NIST\\
NIST  & National Institute of Standards and Technology   \\
NLP & Natural Language Processing\\
PreprocessLM & Language Model used to preprocess input data \\
\hline
\end{tabular}
\end{table}



Technically, \CLLM{} comprises two Language Models: \textit{PreprocessLM}, which is used to convert the diverse fine-tuning and evaluation datasets into a standardized format suitable for machine learning, and \textit{ClassifyLM}, which maps course content to one or multiple Knowledge Areas (KAs). These areas were defined in the CSEC2017 framework and signify key competencies covered in cybersecurity education. ClassifyLM is fine-tuned on the NICE workforce framework, which was labeled by  \cite{ramezanian2024cybersecurity}, and enriched with course-specific knowledge using a synthetic dataset derived from the CSEC2017 framework. 
Several Natural Language Processing (NLP) models were evaluated as candidates for ClassifyLM. A fine-tuned Bidirectional Encoder Representations
from Transformers (BERT) \citep{devlin2019bert} model ultimately outperformed both zero-shot autoregressive models like ChatGPT \citep{ChatGPT} and traditional machine learning baselines. The KAs extracted from ClassifyLM can then be used to match course content to related job roles, as listed in the NICE framework. 


Since the KAs can overlap by design, KA classification is a multi-label task. Furthermore, subjectivity in the definitions of KAs leads to low inter-rater agreement. Therefore, we compare the KAs assigned by \CLLM{} with 3 cybersecurity experts, that have a background in analyzing the CSEC2017 framework, as well as a control group of more than 60 cybersecurity experts inquired at scientific workshops. Our results demonstrate that \CLLM{} can effectively replicate expert judgments, making it the first LLM-based method shown to automate cybersecurity curriculum analysis supported by quantitative and qualitative validation.  

To demonstrate the practical application of \CLLM{}, we present an evaluation of three real-world cybersecurity programs using \CLLM{}: the M.Sc. programs in cybersecurity at Kungliga Tekniska Högskolan (KTH) in Sweden \citep{kth_ms_cybersecurity}, Nanyang Technological University (NTU) in Singapore \citep{ntu_msc_cybersecurity} and the M.Sc. in Information Technology-Information Security (MSIT-IS) offered at Carnegie Mellon University (CMU) in the United States \citep{cmu_msit_is}. Our analysis examines these programs at varying levels of granularity, ranging from individual courses to a program-level analysis. 

To ensure that our methodology reflects the latest workforce requirements, we utilize \CLLM{} to annotate the updated 2025 NICE workforce framework, developed by the National Institute of Standards and Technology \citep{NICE2025}. Specifically, \CLLM{} assigns KAs to each Knowledge Description (KD) within the framework, where KDs represent competency requirements defining knowledge that cybersecurity professionals should possess. The extracted KA labels are benchmarked against expert annotations and subsequently used to update the role-based weightings introduced by \cite{ramezanian2024cybersecurity}. This process ensures that \CLLM{}'s mapping between cybersecurity curricula and job roles reflects the most recent competency requirements. 
Furthermore, we conduct a longitudinal analysis to examine the evolution of these competency demands between 2017 and 2025, and we identify which KAs should be covered in education based on current job market demands. In doing so, this work addresses the persistent cybersecurity workforce skills gap.


Therefore, our key contributions are as follows.

\begin{itemize}
    \item \textbf{A Cybersecurity Curriculum Assessment Tool}: We introduce a novel LLM-based tool called \CLLM{}, designed to assess the compliance of existing cybersecurity curricula with cybersecurity job market needs. This tool is intended for use by 
     a) curriculum developers aiming to align their programs with evolving industry and workforce demands, and 
     b) students or their guidance counselors seeking to tailor students' educational pathways to meet the requirements of specific cybersecurity roles or general job market needs. 
     \item \textbf{Guidelines for LLM-based Curriculum Design}: We present a comprehensive methodology for leveraging fine-tuned large language models (LLMs) in cybersecurity curriculum development. This includes dataset preparation strategies, quantitative comparisons of alternative machine learning approaches, and validation on real-world curricula. This approach can easily be adapted to areas unrelated to cybersecurity. 
     \item \textbf{Workforce-Aligned design}: 
      We extend \CLLM{} to integrate the 2025 NICE Workforce Framework, ensuring that its job role-based weighting scheme remains aligned with the most recent definitions of cybersecurity competencies. 
\end{itemize}

This paper is organized as follows. \secref{relevant_work} provides background on cybersecurity curriculum design and LLMs. \secref{example} presents a case study of personalized curriculum design with \CLLM{}. \secref{methods} details our methodology, including dataset preparation, LLM fine-tuning, and computing job role-based weights based on workforce demands. \secref{results} reports quantitative and qualitative experimental results. \secref{discussion} explores limitations, directions for future research, and alternative applications of \CLLM{}. Finally, \secref{conclusion} summarizes our main findings.

\section{Background} \label{sec:relevant_work}

In this section, we present background knowledge and state-of-the-art literature on cybersecurity curriculum development (Section \ref{sec:lit_cybersec_curr}), automated curriculum planning (Section \ref{sec:autom_curr}), and LLMs (Section \ref{sec:LLMs}). 

\subsection{Cybersecurity Curriculum Development}\label{sec:lit_cybersec_curr} 
The significance of developing, maintaining, and continuously improving cybersecurity curricula has been widely acknowledged within the research community \citep{ismail2024cybersecurity}. Consequently, numerous scholarly works have addressed on this topic from various perspective, focusing on aspects such as methods on enhancing cybersecurity pedagogy \citep{pirta2024try, barbosa2024cybersecurity}, workforce training on both national and international levels \citep{pirta2025latvian, davis2025best,  ramezanian2024cybersecurity}, developing quality criteria for cyber security MOOCs (Massive Open Online Courses) \citep{fischer2020quality}, and the identification of essential topics to be included in cybersecurity curricula \citep{barbosa2024education, aranzazu2025cyber}. 

This study builds upon the following foundational frameworks.
\begin{itemize}
    \item The \textit{Cybersecurity Curricula 2017 (CSEC2017)} framework, developed by the Joint Task Force on Cybersecurity Education (consisting of professionals from ACM, IEEE, AIES, and IFIP) \citep{csec2017}. CSEC2017 organizes key concepts in the cybersecurity domain into eight Knowledge Areas, which are further decomposed into Knowledge Units. A detailed definition of each KA is presented in \tabref{KAs}.
    
    \item The \textit{National Initiative for Cybersecurity Education (NICE)} workforce framework v2.0.0., published by the National Institute of Standards and Technology \citep{NICE2025}. The NICE framework establishes a standardized taxonomy for the cybersecurity workforce, consisting of 5 workforce categories and 41 cybersecurity work roles. Each work role is decomposed into detailed Knowledge, Skill and Ability descriptions.
    \item The work of \textit{\cite{ramezanian2024cybersecurity}}, which provides a mapping between the NICE and CSEC2017 framework. Their methodology involved manually classifying NICE Knowledge Description into the eight CSEC2017 Knowledge Areas. This process revealed a range of topics relevant to cybersecurity education but not covered by CSEC2017. To address this, they proposed a new, ninth Knowledge Area `Miscellaneous', which includes topics such as Computer Science and Information Technology. 
    Note that this work was based on an earlier version of the NICE framework (v1.0.0. \citep{NICE2017}). For an overview of the changes between the two versions of the NICE framework, we refer the reader to Section \ref{sec:job_to_KA}. 
    

\end{itemize}


\newcommand{\bigbullet}[1]{\resizebox{!}{.25cm}{$\textcolor{#1}\bullet$}}
\begin{table}[htb]
    \centering
\resizebox{.98\textwidth}{!}{
    \begin{tabular}{cclm{9cm}}
    \toprule
         Color & Number & Knowledge Area (KA) & Concise Description  \\
    \midrule
         \bigbullet{KA6} & 0 & Miscellaneous & Assorted topics: Computer Science, Business and Law, Communication and Networking, Information Technology, Cyberspace Practice, Pedagogy, Intelligence\\
\graybg  \bigbullet{KA9} & 1 & Data Security & Basic cryptography concepts, Digital forensics, End-to-end secure communications, Data integrity and authentication, Information storage security\\
         \bigbullet{KA1} & 2 & Software Security & Fundamental design principles, Security requirements, Implementation issues, Static and dynamic testing, Configuring and patching, Ethics\\
\graybg  \bigbullet{KA2} & 3 & Component Security & Vulnerabilities of system components, Component lifecycle, Secure component design principles, Supply chain management security, Security testing, Reverse engineering\\
         \bigbullet{KA3} & 4 & Connection Security & Systems, architecture, models, standards, Physical component interfaces, Software component interfaces, Connection attacks, Transmission attacks\\
\graybg  \bigbullet{KA4} & 5 & System Security & Holistic approach, Security policy, Authentication, Access control, Monitoring, Recovery, Testing, Documentation\\
         \bigbullet{KA5} & 6 & Human Security & Identity management, Social engineering, Awareness and understanding, Social behavioral privacy and security, Personal data privacy and security\\
\graybg  \bigbullet{KA7} & 7 & Organizational Security & Risk management, Governance and policy, Laws, ethics, and compliance, Strategy and planning\\
         \bigbullet{KA8} & 8 & Societal Security & Cybercrime, Cyber law, Cyber ethics, Cyber policy, Privacy\\
    \bottomrule
    \end{tabular}
    }
    \caption{Classification of Cybersecurity Concepts. This table outlines the eight CSEC2017 Knowledge Areas \citep{csec2017} (corresponding to numbers 1--8) and the Miscellaneous class (corresponding to 0) as introduced by \cite{ramezanian2024cybersecurity} and detailed in Appendix \ref{KnowledgeArea0:Miscellaneous}. The color coding and numbering of KAs is used consistently throughout the paper to enhance readability.}
    \label{tab:KAs}
\end{table}

\subsection{Automated Curriculum Planning}
\label{sec:autom_curr}

The first reported use of AI for curriculum planning was by \cite{evans1976automated}. However, this concept received limited attention until recent years, when AI began to play a more prominent role in educational applications such as formulating learning objectives \citep{sridhar2023harnessing}, intelligent tutoring systems, and automated assessment \citep{harry2023role}. 

Classical approaches to automate curriculum design include heuristic-based methods, such as course dependency graphs \citep{backenkohler2018data} and genetic algorithms \citep{wong2018sequence}. More recently, deep learning frameworks employing Long Short Term Memory (LSTM) networks, a type of  Recurrent Neural Networks, have been applied to curriculum design \citep{duan2019automatic, dzurenda2024enhancing}. The advent of Large Language Models (LLMs) has further advanced the field by enabling fully automated curriculum design, with \cite{dzurenda2024enhancing} representing the first work to apply LLMs specifically to cybersecurity curriculum development. This work serves as a key baseline for \CLLM{}. 

\subsection{LLMs}
\label{sec:LLMs}
Natural Language Processing (NLP) focuses on the machine learning-based processing of textual data. 
Since neural networks are designed to process numerical inputs rather than text, all NLP methods first convert textual data into numerical representations through a process known as tokenization. Historically, this was achieved using frequency-based feature extraction techniques such as bag-of-words (BoW) \citep{mctear2016conversational} or frequency-inverse document frequency (TF-IDF) \citep{sparck1972statistical}, which assign a number to each unique word in a document based on its frequency of occurrence.

Nowadays, deep learning methods for NLP or so-called Large Language Models (LLMs) have revolutionized the field of NLP. These models map words to numerical tokens via \textit{training} on large-scale text corpora \citep{devlin2019bert}. 
Unlike traditional NLP methods that break up sentences into unordered sets of words, LLMs preserve and leverage the sentence structure and context.  

LLMs can be categorized into autoencoding and autoregressive models based on their architectural structure. 
Autoencoding models, such as Bidirectional Encoder Representations from Transformers (BERT), are trained using a denoising objective where inputs are masked and an encoder-decoder model is optimized to recover the unmasked input. BERT uses bidirectional self-attention to incorporate context from both preceding and subsequent words within a sentence.
In contrast, autoregressive models such as the GPT models (including ChatGPT \citep{ChatGPT}), are trained to predict the next word in a sentence given the preceding context. 

While off-the-shelf LLMs perform well on general-purpose tasks, domain adaptation is often necessary for specialized applications, such as curriculum design. In the case of autoencoding models, this can be done by fine-tuning the parameters of an off-the-shelf model on a small domain-specific dataset. Autoregressive models, in contrast, are typically adapted through few-shot learning on a limited number of examples, allowing modification of the model's behavior without explicitly altering the underlying parameters \citep{zhou2024comprehensive}. 
In this study, we fine-tune BERT \citep{devlin2019bert}, an autoencoding model, that comprises approximately 110 million parameters. BERT is particularly suitable for text classification tasks \citep{qiu2024chatgpt}, e.g. classifying course content into one of the nine KAs (eight KAs from  CSEC2017 and one KA from \cite{ramezanian2024cybersecurity}). Additionally, our training dataset is relatively small, which makes fine-tuning larger models infeasible in practice due to both the substantial computational cost and the increased risk of overfitting when adapting models with billions of parameters. For context, ChatGPT-4o mini is estimated to contain around 8 billion parameters \citep{paramsGPT}, highlighting the significant computational disparity between the two approaches.



To date, \cite{dzurenda2024enhancing} present the only other LLM-assisted curriculum design tool. Notably, their methodology differs fundamentally from ours: although they utilize the BERT \textit{tokenizer}, they substitute the subsequent BERT \textit{base} and \textit{classification} model with a Long Short-Term Memory (LSTM) model. Furthermore, their model is fine-tuned based on job advertisements, whereas our approach explicitly fine-tunes on educational standards and maps curricula to workforce requirements. In our fine-tuning methodology (Section \ref{sec:LL2_finetuning} and \ref{app:finetuning}), we adopt the methodology proposed by \cite{dzurenda2024enhancing} as a baseline to evaluate the performance of our approach.

\section{Case Example: Personalized Curriculum Design using \CLLM{}} \label{sec:example}

To illustrate how \CLLM{} can be used for personalized curriculum design, we analyze the following case example: 
\textit{Alice is a Master's student in cybersecurity at KTH. They want to become a vulnerability analyst. Which elective courses should they take to best prepare themselves for this role?}

\begin{figure}[htbp!]
\centering
\resizebox{.95\textwidth}{!}{
\begin{tikzpicture}
[
    node distance=0.8cm,
    program/.style={rectangle, draw, rounded corners, fill=blue!10, minimum width=3cm, minimum height=1.2cm, text width=2.8cm, align=center},
    elective/.style={rectangle, draw, rounded corners, fill=green!10, minimum width=2.7cm, minimum height=0.7cm, text width=2.5cm, align=center},
    selected/.style={rectangle, draw, rounded corners, fill=green!20, minimum width=3cm, minimum height=1.2cm, text width=2.8cm, align=center},
    pie/.style={circle, draw=black, minimum width=1cm},
    legend/.style={font=\footnotesize, anchor=west},
    arrow/.style={->, >=stealth, thick},
    mydashed/.style={dashed, ->, >=stealth, thick}
]
{

\node[program] (electives) {Elective Courses \\ (7.5 ECTS each)};
\node[program, below right=3.2cm and 3.5cm of electives] (select) {Select 4 of 12 \\ Electives};

\node[elective, below left=0.5cm and 0cm of electives] (e1) {Foundations of \\ Cryptography};
\node[right=0.2cm of e1] (e1pie) {
\begin{tikzpicture}[baseline=-0.5ex]
    \draw[fill=KA9] (0,0) circle (0.4cm);
\end{tikzpicture}};
\node[draw, rounded corners,fit=(e1) (e1pie)] {};
\draw[arrow] (e1) -- (e1pie);

\node[elective, right=1.6cm of e1] (e2) {Priv. Enhancing \\ Technology};

\node[right=0.2cm of e2] (e2pie) {
    \piechart[0.4]{
    21.4/KA9,
    21.4/KA5, 
    28.6/KA7, 
    28.6/KA8}{}
};

\node[draw, rounded corners,fit=(e2) (e2pie)] {};
\draw[arrow] (e2) -- (e2pie);

\node[elective, below=0.5cm of e1] (e3) {Project in \\ System Sec.};

\node[right=0.2cm of e3] (e3pie) {
    \piechart[0.4]{
    8.7/KA6, 
    30.4/KA1, 
    26/KA2, 
    8.7/KA3, 
    8.7/KA4, 
    17.4/KA7}{}
};

\node[draw, rounded corners,fit=(e3) (e3pie)] {};
\draw[arrow] (e3) -- (e3pie);

\node[elective, right=1.6cm of e3] (e4) {Language-based \\ Security};
\node[right=0.2cm of e4] (e4pie) {
    \piechart[0.4]{
    44.4/KA1, 
    22.2/KA5, 
    33.3/KA7}{}
};
\node[draw, rounded corners,fit=(e4) (e4pie)] {};
\draw[arrow] (e4) -- (e4pie);

\node[elective, below=0.5cm of e3] (e5) {Cyber-Physical \\ Security };
\node[right=0.2cm of e5] (e5pie){
    \piechart[0.4]{
    4/KA6, 
    36/KA9,
    4/KA1, 
    4/KA2, 
    12/KA3, 
    16/KA4, 
    20/KA7, 
    4/KA8}{}
};
\node[draw, rounded corners,fit=(e5) (e5pie)] {};
\draw[arrow] (e5) -- (e5pie);

\node[elective, right=1.6cm of e5] (e6) { Networked \\ Systems Sec.};
\node[right=0.2cm of e6] (e6pie) {
    \piechart[0.4]{
    12.5/KA6, 
    50/KA3, 
    37.5/KA4}{}
};
\node[draw, rounded corners,fit=(e6) (e6pie)] (e6box) {};
\draw[arrow] (e6) -- (e6pie);

\node[elective, below=0.5cm of e5] (e7) {Advanced Netw.\\ Systems Sec.};
\node[right=0.2cm of e7] (e7pie) {
    \piechart[0.4]{
    10/KA6, 
    20/KA1, 
    40/KA3, 
    20/KA4, 
    10/KA7}{}
};
\node[draw, rounded corners,fit=(e7) (e7pie)] {};
\draw[arrow] (e7) -- (e7pie);

\node[elective, right=1.6cm of e7] (e8) {Building Netw.\\ Systems Sec. };
\node[right=0.2cm of e8] (e8pie) {
\piechart[0.4]{
    12.5/KA6, 
    12.5/KA3,  
    12.5/KA4, 
    62.5/KA7}{}
};

\node[draw, red, thick, rounded corners,fit=(e8) (e8pie)] (e6box) {};
\draw[arrow] (e8) -- (e8pie);

\node[elective, below=0.5cm of e7] (e9) {Digital\\ Forensics};
\node[right=0.2cm of e9] (e9pie){
    \piechart[0.4]{
    88.9/KA9,
    11.1/KA7}{}
};
\node[draw, rounded corners,fit=(e9) (e9pie)] {};
\draw[arrow] (e9) -- (e9pie);

\node[elective, right=1.6cm of e9] (e10) {Sec. Analysis \\ Large-Scale Sys.};
\node[right=0.2cm of e10] (e10pie) {
    \piechart[0.4]{
    22.2/KA6, 
    11.1/KA3, 
    22.2/KA4, 
    44.4/KA7}{}
};
\node[draw, rounded corners,fit=(e10) (e10pie)] {};
\draw[arrow] (e10) -- (e10pie);

\node[elective, below=0.5cm of e9] (e11) {Design of\\ Fault-tol. sys.};
\node[right=0.2cm of e11] (e11pie) {
    \piechart[0.4]{
    8.7/KA6, 
    8.7/KA9,
    34.8/KA1, 
    8.7/KA2, 
    8.7/KA3, 
    13.0/KA4,  
    17.4/KA7}{}
};

\node[draw, rounded corners,fit=(e11) (e11pie)] (box11) {};
\draw[arrow] (e11) -- (e11pie);

\node[elective, right=1.6cm of e11] (e12) {Hardware\\  Security};
\node[right=0.2cm of e12] (e12pie) {
    \piechart[0.4]{ 
    10/KA9,
    10/KA1, 
    20/KA4, 
    30/KA5, 
    10/KA7, 
    20/KA8}{}
};
\node[draw, rounded corners,fit=(e12) (e12pie)] (box12) {};
\draw[arrow] (e12) -- (e12pie);

\node[program, right=3.5cm of electives] (core) {Mandatory Courses \\ (39.5 ECTS)};
\node[below=0.2cm of core] (corepie) {
    \piechart[0.5]{
    7.7/KA6, 
    18.4/KA9,
    4.9/KA1, 
    4.9/KA2, 
    5.2/KA3, 
    12.5/KA4, 
    5.3/KA5, 
    24.2/KA7, 
    17.0/KA8}{}
};


\node[draw, dashed, rounded corners,fit=(core) (corepie)] (all_core) {};

\node[draw, dashed, rounded corners,fit=(electives) (box12) (e11) (box11) (e12)] (all_electives) {};

\node[above right=-3.5cm and 5cm of electives, xshift=2.5cm, align=left, font=\footnotesize] {
      \textbf{Security KA:} \\
   $\color{KA6}\bullet$ Miscellaneous\\
   $\color{KA9}\bullet$ Data\\
    $\color{KA1}\bullet$ Software\\
    $\color{KA2}\bullet$ Component\\
    $\color{KA3}\bullet$ Connection\\
    $\color{KA4}\bullet$ System\\
    $\color{KA5}\bullet$ Human\\
    $\color{KA7}\bullet$ Organizational\\
    $\color{KA8}\bullet$ Societal
    };

\node[program, right= 1.2cm of select](nodee){Final program \\ (69.5 ECTS)};

\node[below=0.2cm of nodee] (selpie) {
    \piechart[0.5]{
    6.8/KA6, 
    12.8/KA9,
    4.9/KA1, 
    2.8/KA2, 
    15.4/KA3, 
    14.6/KA4, 
    5.3/KA5, 
    24.7/KA7, 
    12.8/KA8}{}
};

\node[below=1.7cm of nodee, xshift=0cm, yshift=-0cm, align=left, font=\footnotesize] (selected) {
\textbf{Selected Courses: } \\
$\bullet$ Networked System Sec. \\
$\bullet$ Adv. Netw. Systems Sec.  \\
$\bullet$ Building Netw. Systems Sec. \\
$\bullet$ Priv. Enhancing Tech.  
};
\node[draw, dashed, rounded corners,fit=(selected) (nodee)] (node2) {};

\node[below= 1.75cm of selected, xshift=-6.5cm] (fig2) {
\begin{tikzpicture}[
    node distance=5mm,
    title/.style={font=\bfseries,align=center},
    process/.style={rectangle, draw, minimum width=2.5cm, minimum height=1cm, text width=3cm, align=center},
    result/.style={rectangle, draw, minimum width=2.5cm, minimum height=1cm, text width=2.5cm, align=center},
    note/.style={rectangle, draw=black!50, fill=black!10, thick, minimum height=1cm, align=left},
    arrow/.style={->, >=stealth, thick}
]


\node[process, xshift=0cm, yshift=0.1cm] (step1) {Course Title \& Course Description};
\node[trapezium, trapezium angle=80, rotate=90, draw=blue!50, fill=blue!20, minimum width=27mm, draw, below right=-0.1cm and 2cm of step1] (step2) {
\begin{tikzpicture}
    \node [rotate=-90, text width=3.2cm] (step2a) {Extract subtopics \\  using \textbf{PreprocessLM}};
\end{tikzpicture}};

\node[trapezium, trapezium angle=80, rotate=90, draw=green!50, fill=green!20, minimum width=27mm, draw, below right=1.85cm and 2.5cm of step2] (step3) {
\begin{tikzpicture}
    \node [rotate=-90, text width=3.2cm] (step3a) {Classify into KAs \\  using \textbf{ClassifyLM}};
\end{tikzpicture}};

\node[result, below right=0.6 and 1.5cm of step3] (final) {Processed Course Data};

\draw[arrow] (step1) -- (step2);
\draw[arrow] (step2) -- (step3);
\draw[arrow] (step3) -- (final);

\node[note, below=of step1, xshift=0cm, text width=5.5cm] (example) {
    \textbf{Example Course:} Building Networked Systems Security (EP2520)\\
    \textit{Description:} The course trains students to handle contemporary security problems for networked systems. Through problem-based learning and team-work, students tackle modern technical challenges for cybersecurity. Students investigate requirements, design specifications, prepare solutions with professional tools and critically assess the efficiency of alternative solutions. 
};

\node[note, below=1.13cm of step2, yshift=-0.3cm, xshift=0.5cm, text width=8cm] (topics) {
    \textbf{Extracted Topics:} \\
    $\bullet$ building networked systems security \\
    $\bullet$ handling contemporary sec. problems for netw. sys. \\
    $\bullet$ problem-based learning \\
    $\bullet$ teamwork in cybersecurity \\
    $\bullet$ investigating requirements for netw. sys. sec. \\
    $\bullet$ designing specifications for cybersecurity \\
    $\bullet$ preparing solutions with professional tools \\
    $\bullet$ critically assessing the efficiency of alternative solutions \\
};
\node[note, below=1.35cm of step3, xshift=1cm, yshift=-0.3cm, text width=1.3cm] (classification) {
    \textbf{KAs:} \\
    $\bullet$ 7 \\
    $\bullet$ 4 \\
      $\bullet$ 0 \\
    $\bullet$ 7 \\
    $\bullet$ 5 \\
    $\bullet$ 7\\
    $\bullet$ 7\\
    $\bullet$ 7\\
};

\node[note, below=0.85cm of final, xshift=0.6cm, yshift=-0.3cm, text width=1.3cm] (KAs) {
    \textbf{KAs:} \\
\piechart[0.5]{
    12.5/KA6, 
    12.5/KA3,  
    12.5/KA4, 
    62.5/KA7}{}
};

\draw[dashed] (step1) -- (example);
\draw[dashed] (step2) -- (topics);
\draw[dashed] (step3) -- (classification);
\draw[dashed] (final) -- (KAs);
\end{tikzpicture}
};

\node[draw, red, fit=(fig2)] (fig2box) {};
\draw[dashed,red] (e6box.north east) -- (fig2box.north east);
\draw[dashed,red] (e6box.north west) -- (fig2box.north west);


\node[program, below=6cm of core] (job) {Vulnerability Analyst \\ job requirements};
\node[below=0.2cm of job] (jobpie) {
    \piechart[0.5]{
    9/KA6, 
    10/KA9,
    4/KA1,  
    14/KA3, 
    6/KA4, 
    9/KA5, 
    36/KA7, 
    14/KA8}{}};

\node[draw, dashed, rounded corners,fit=(job) (jobpie)] (all_job) {};
\draw[arrow] (all_electives) -- (select);
\draw[arrow] (all_core) -- (select);
\draw[arrow] (select) -- (nodee);
\draw[arrow] (all_job) -- (select);
};
\end{tikzpicture}
}

\caption{\CLLM{} applied to the KTH MSc in Cybersecurity (Case Study in \secref{example}). Knowledge Area distributions are indicated as pie charts, following the color scheme as displayed in the legend on the top right of the figure. Each possible recommended elective is shown on the top left, the aggregated KA distribution of the mandatory part of the curriculum is shown in the center, as well as the required KA distribution for the Vulnerability Analyst job role. The KA labeling process is illustrated in greater detail for the Building Networked Systems Security course on the bottom of the figure. The selected elective courses and the final KA composition of the program are listed on the right. }
\label{fig:example}
\end{figure}
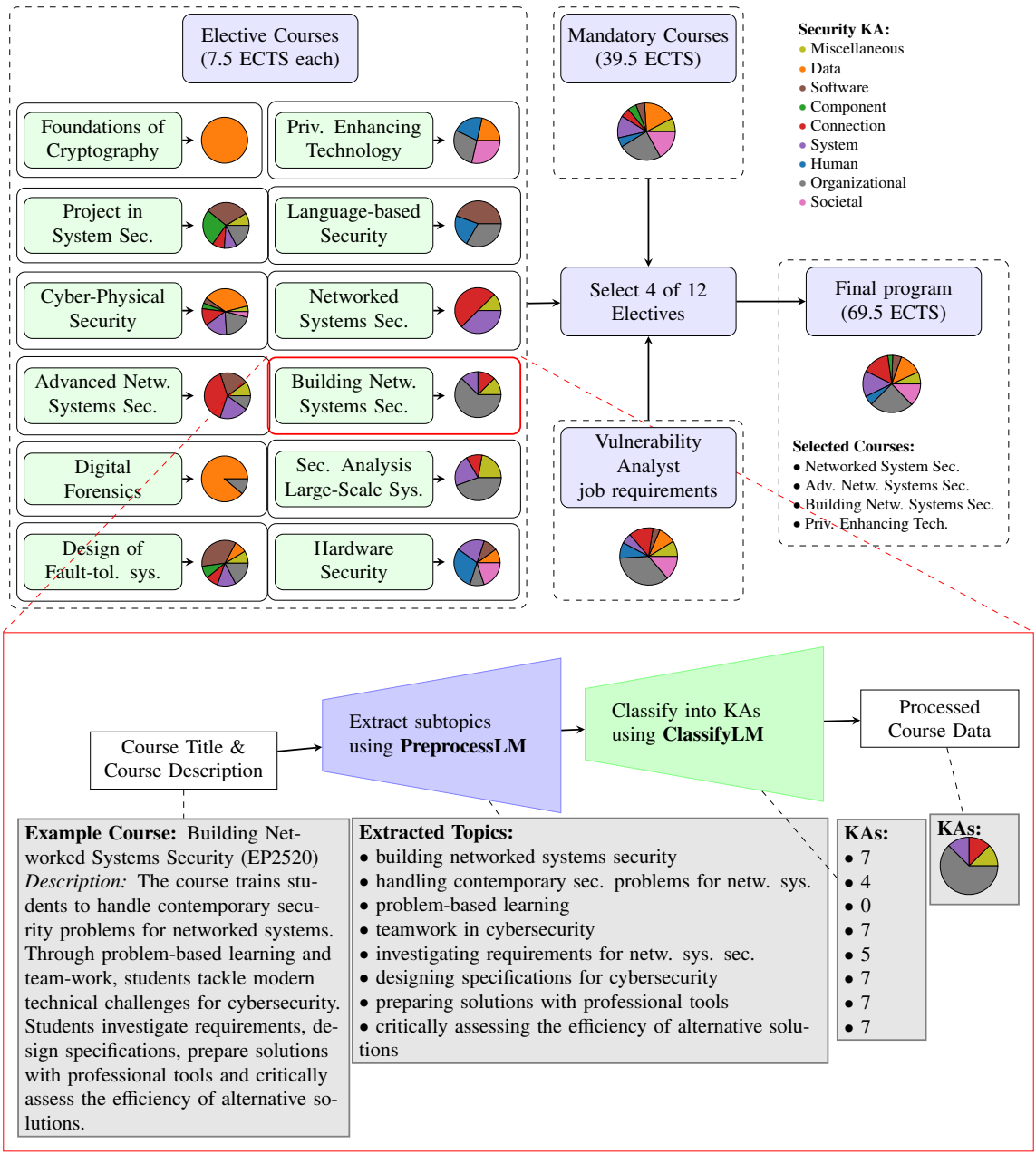

We address this problem methodologically, as illustrated in \figref{example}. 
Alice must select 4 out of the 12 elective courses listed on the top left of \figref{example}. 
In this work, we adopt the Knowledge Areas, as proposed in the CSEC2017 curriculum \citep{csec2017}, to compare academic courses with job market demands. A brief description of all KAs is provided in \tabref{KAs}. Each KA encompasses an essential component of cybersecurity education, with the exception of the `Miscellaneous' KA, which covers topics previously not included in the CSEC2017 KAs \citep{ramezanian2024cybersecurity}. The color-coding scheme used throughout this paper is also highlighted on the top right of \figref{example}.

For each mandatory and elective course in the KTH MSc in Cybersecurity we apply \CLLM{} to compute the distribution of KAs covered in the course. A representative example, the Building Networks System Security elective course, is depicted at the bottom of \figref{example}. The KA labeling process follows three stages: 

\begin{enumerate}
    \item \textbf{Course preprocessing}  We extract the course title and course description, which are then fed through an instruction-tuned Large Language Model (PreprocessLM) to generate concise topic descriptors that capture the core content of the course. 
    \item \textbf{KA classification} Each extracted topic is subsequently passed through a multi-label classification Language Model (ClassifyLM). ClassifyLM was fine-tuned on cybersecurity topics derived from the CSEC2017 curriculum and Knowledge Descriptions extracted from the NICE workforce framework, to be able handle both course and workforce data. 
    \item \textbf{Label aggregation} Finally, all KA labels extracted from the course title and description are aggregated, normalized, and visualized as a pie chart. 
\end{enumerate}

The Building Networked System Security course focuses on training students to handle contemporary security problems through problem-based learning and teamwork. 
As expected, the extracted topic `handling contemporary security problems for networked systems' was assigned KA 4, Connection Security. The aggregated KA labels show that Organizational Security dominates, and System Security, Connection Security, and Miscellaneous account for the remainder of the course content. 



This procedure is systematically applied to all courses in the program (see the pie charts for each elective in \figref{example}). The aggregated KA distribution for the mandatory component of the Cybersecurity master at KTH is displayed in the center top of \figref{example}. Simultaneously, we map the KA profile for the target role of vulnerability analyst based on the mappings provided in Figure \ref{fig:job_role_KAs_combined}. 
When comparing the KA distribution of Alice's mandatory courses and the KA distribution of the vulnerability analyst job role, we see that Alice primarily lacks courses related to Connection Security (in red) and Miscellaneous (in lime). 
To identify the most suitable set of four elective courses to address this knowledge gap, we formulate a constrained optimization problem (for more details on this formulation, see \ref{sec:constr_optimization}). The objective is to minimize the divergence between the combined KA distribution of the selected curriculum (mandatory \& selected electives) and that of the target job role.

In Alice's case, the optimization identifies four electives that most closely align their curriculum with the desired role of vulnerability analyst (see the middle-right of \figref{example}).
The Networked Systems Security course and the Advanced Networked Systems Security course contain the largest proportion of content related to Connection Security, while Building Networked Systems Security includes the largest proportion of Organizational Security out of all elective courses. The resulting final program aligns more closely to the required KA distribution for the Vulnerability Analyst job role than the stand-alone mandatory program. Therefore, Alice's electives have been chosen in such a way that they increase the correspondence of their curriculum with the requirements from the job market.



\section{Methodology} \label{sec:methods}

In this work, we present \CLLM{}, a novel end-to-end framework that leverages LLMs for the analysis of cybersecurity curricula. \CLLM{} consists of the following steps: data preprocessing, KA annotation, and the computation of job role-based weights based on demands in the cybersecurity workforce. 
The remainder of this section provides a comprehensive description of each component, beginning with data selection and pre-processing (\secref{datasets}), followed by the fine-tuning procedure for ClassifyLM (\secref{LL2_finetuning}), and the computation of job role-based weights (\secref{role-basedweight}). 

\subsection{Datasets} \label{sec:datasets}

\begin{table*}[htb!]
    \centering
\resizebox{.98\textwidth}{!}{
    \begin{tabular}{p{1.6cm}p{5cm}cccp{4cm}}
    \toprule
   Name & Example &  Purpose & PreprocessLM & Labelled & Example after preprocessing \\
    \midrule

    NIST KDs & \textit{K0018: Knowledge of encryption algorithms} & Fine-tuning & \xmark & \cmark / \xmark &  $[1]$, \textit{encryption algorithms}\\[6mm]
    
 \graybg CSEC2017 synthetic & \textit{KA: 1 \& 4 \newline KU: Software testing \newline Description: This topic describes how to test the software as a whole, and place unit and integration testing in a proper framework}  & Fine-tuning & \cmark & \cmark & \textit{$[1,4]$, software testing \newline
$[1,4 ]$, unit testing \newline $[1,4 ]$, integration testing}\\[24mm]

    KTH, NTU and CMU curricula & \textit{Title: Networked System Security (EP2500) \newline
    Description:  Basically, the course will bring up security including integrity for a spectrum of network systems that includes: Internet and TCP/IP networks, Mobile voice and data networks, Wireless local and  personal networks, Wireless sensor networks, Mobile ad hoc and hybrid networks, such as vehicle communication systems. The emphasis in the  course lies on basic concepts and technologies about joint security requirements in different systems and about how the functions in each system decide the latest security solutions.} & Evaluation & \cmark & \xmark & \textit{$\bullet$ networked system security \newline
    $\bullet$ internet and TCP/IP networks \newline
    $\bullet$ mobile voice and data networks \newline
    $\bullet$ wireless local and personal networks \newline
    $\bullet$ wireless sensor networks \newline
    $\bullet$ mobile ad hoc and hybrid networks \newline
    $\bullet$ vehicle communication systems}\\ 
\bottomrule   
    \end{tabular}
}
    \caption{An overview of all data used to fine-tune and validate \CLLM{}. Columns denote the dataset name, an example entry, the purpose (fine-tuning/evaluation), whether PreprocessLM was applied, whether KA labels were available, and the example after preprocessing, respectively. }
    \label{tab:Data}
\end{table*}

\tabref{Data} presents an overview of all datasets used to fine-tune and evaluate \CLLM{}, along with their respective preprocessing methods. 
The NICE Workforce Framework defines standardized competency specifications for cybersecurity professionals, called Knowledge Descriptions. The KDs contained in NICE framework v1.0.0. were annotated with Knowledge Area labels by \cite{ramezanian2024cybersecurity} and can therefore be used to fine-tune the classification model, ClassifyLM. 
Because the NICE framework focuses on workforce roles rather than educational content, we complement it with course-based data for fine-tuning and evaluation to ensure that the resulting methodology is well aligned with curriculum analysis. This dataset is synthetic and derived from the CSEC2017 framework. This framework comprises nine Knowledge Areas (including the recently proposed Miscellaneous KA \citep{ramezanian2024cybersecurity}), which are further subdivided into ‘Knowledge Units’ (KUs) containing detailed descriptions of cybersecurity topics covered in the KU. PreprocessLM was used to extract fine-grained topics from these descriptions (as shown in \figref{llm-prompt}). The resulting synthetic dataset contains approximately 2,100 labeled topics.

For our evaluation pipeline, we curated a set of real-world course descriptions collected from the official MSc Cybersecurity curricula at the KTH royal academy of Stockholm, Nanyang Technical University of Technology (NTU) in Singapore and Carnegie Mellon University in Pittsburgh. The courses contained in these curricula provide a diverse and realistic benchmark for evaluating to what extent \CLLM{} generalizes to real academic curricula. Furthermore, NIST has recently released an updated version of the NICE framework, in which only 15 Knowledge Descriptions remain unchanged from the previous release \citep{NICE2025}. This framework therefore also used to evaluate our methodology.

LLMs are known to be sensitive to stylistic features such as the prompt length \citep{levy2024same}, uppercase characters \citep{vivi2025uppercase}, and the formatting of prompts as plaintext, markdown, YAML or JSON \citep{he2024does}. 
Therefore, our preprocessing pipeline is designed to uniformize the three datasets in \tabref{Data}. We convert all data into a standardized format consisting of a single sentence that describes one cybersecurity topic, devoid of unnecessary verbosity. For the NIST KDs, this entailed removing the redundant prefix ``Knowledge of'', while the other datasets were processed more rigorously with PreprocessLM.  
As illustrated in \tabref{Data}, the datasets are more homogeneous after pre-processing: All entries are short, in lower-case, and redundant text has been removed. This is particularly noticeable for the CSEC2017 synthetic and KTH, NTU and CMU curricula data. The data standardization process ensures that both fine-tuning and evaluation data follow similar formatting structures, enhancing model robustness.

\begin{figure}[ht]
\centering
\begin{tcolorbox}[colback=gray!5!white,colframe=black!75!white,
  title=Prompt Structure for Subtopic Extraction Using PreprocessLM]
\textbf{System Prompt:}\\
\texttt{
"You are a helpful AI assistant. Instructions: \\
a. Carefully read the topic and the description.\\
b. Provide a list of all subtopics contained within the description.\\
c. Do not include any explanation or additional text in the response."}

\vspace{0.5em}
\textbf{User Message:}\\
\texttt{
"\#topic: \{topic\}, \#description: \{description\}"}
\end{tcolorbox}
\caption{The prompt template used for PreprocessLM to extract subtopics from a larger piece of text. The placeholders \{topic\} and \{description\} are replaced by the Knowledge Unit (or course title) and its corresponding detailed description, respectively. }
\label{fig:llm-prompt}
\end{figure}

\subsection{ClassifyLM fine-tuning} \label{sec:LL2_finetuning}

We evaluated various different modeling strategies to effectively classify course descriptions into one or multiple Knowledge Areas, ranging from traditional machine learning methods to fine-tuned LLMs. The fine-tuned BERT model \citep{devlin2019bert} emerged as the best candidate to serve as \textit{ClassifyLM}, due to its superior KA classification performance on the fine-tuning dataset (described in Section \ref{sec:datasets}), surpassing traditional machine learning techniques, the baseline developed by \cite{dzurenda2024enhancing}, and zero-shot prompting of ChatGPT-4o-mini \citep{ChatGPT} and DeepSeek-V3 \citep{liu2024deepseek}. The full fine-tuning and model selection methodology is outlined in \ref{app:finetuning}.
The final ClassifyLM model was trained using the following settings: 
\begin{itemize} 
\item Model: BERT (listed as bert-base-uncased in the HuggingFace library)
\item Loss Function: Binary Cross-Entropy
\item Optimizer: Adam
\item Batch Size: 64
\item Learning rate: $4\times 10^{-5}$
\item Epochs: 10
\end{itemize}
These hyperparameters were selected based on cross-validated macro-F1 scores. 
All code containing baselines, our implementation of \CLLM{}, data preprocessing and evaluation methodologies has been included on our publicly available GitHub \citep{github_CLLM}. 

\subsection{Cybersecurity job role-based weights} \label{sec:role-basedweight}

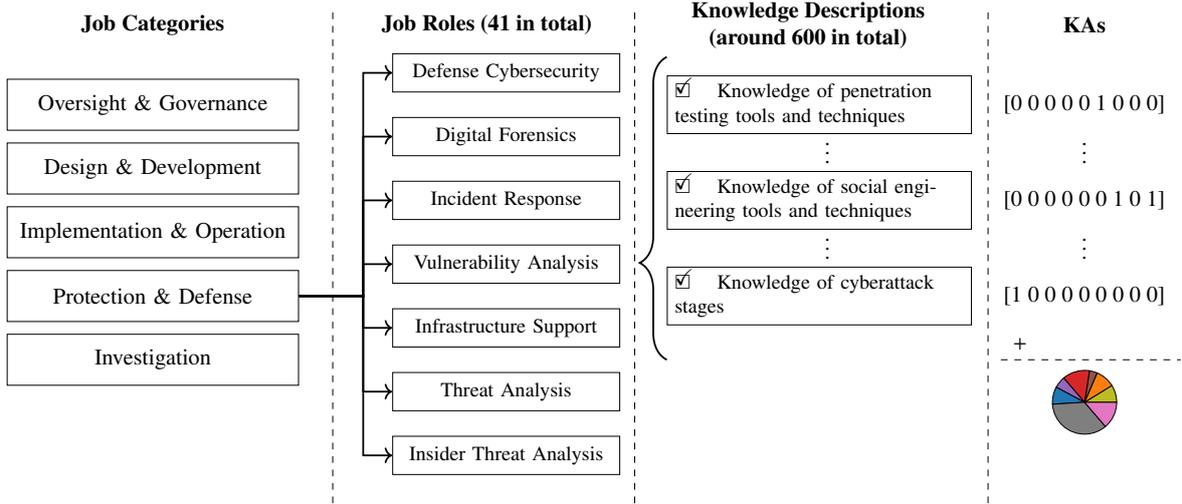
\begin{figure}[htbp]
\centering
\resizebox{.98\textwidth}{!}{
\begin{tikzpicture}[
    box/.style={draw, rectangle, text width=4.3cm, minimum height=0.8cm, align=center},
    rolebox/.style={draw, rectangle, text width=3.3cm, minimum height=0.6cm, align=center, font=\small},
    knowledgebox/.style={draw, rectangle, text width=4.5cm, minimum height=0.5cm, align=left, font=\small, anchor=west},
    title/.style={font=\large\bfseries, align=center},
    section/.style={font=\bfseries, align=center}
]

\node[section] at (-5,3.75) {Job Categories};
\node[section] at (0.2,3.75) {Job Roles (41 in total)};
\node[section] at (5.2,3.75) {Knowledge Descriptions \\ (around 600 in total)};
\node[section] at (9.5,3.75) {KAs};

\node[box] (og) at (-5,2.5) {Oversight \& Governance};
\node[box] (dd) at (-5,1.5) {Design \& Development};
\node[box] (io) at (-5,0.5) {Implementation \& Operation};
\node[box] (pd) at (-5,-0.5) {Protection \& Defense};
\node[box] (in) at (-5,-1.5) {Investigation};

\node[rolebox] (defense) at (0.5,3) {Defense Cybersecurity};
\node[rolebox] (forensics) at (0.5,2) {Digital Forensics};
\node[rolebox] (incident) at (0.5,1) {Incident Response};
\node[rolebox] (infrastructure) at (0.5,0) {Vulnerability Analysis};
\node[rolebox] (insider) at (0.5,-1) {Infrastructure Support};
\node[rolebox] (threat) at (0.5,-2) {Threat Analysis};
\node[rolebox] (vulnerability) at (0.5,-3) {Insider Threat Analysis};

\draw[->, thick] (pd.east) -- ($(pd.east)+(1,0)$) |- (defense.west);
\draw[->, thick] (pd.east) -- ($(pd.east)+(1,0)$) |- (forensics.west);
\draw[->, thick] (pd.east) -- ($(pd.east)+(1,0)$) |- (incident.west);
\draw[->, thick] (pd.east) -- ($(pd.east)+(1,0)$) |- (infrastructure.west);
\draw[->, thick] (pd.east) -- ($(pd.east)+(1,0)$) |- (insider.west);
\draw[->, thick] (pd.east) -- ($(pd.east)+(1,0)$) |- (threat.west);
\draw[->, thick] (pd.east) -- ($(pd.east)+(1,0)$) |- (vulnerability.west);

\node[knowledgebox] (k1) at (3,2.5) {$\square$ \hspace{0.2cm} Knowledge of penetration testing tools and techniques};
\node[knowledgebox] (k2) at (3,1) {$\square$ \hspace{0.2cm} Knowledge of social engineering tools and techniques};
\node[knowledgebox] (k3) at (3,-0.5) {$\square$ \hspace{0.2cm} Knowledge of cyberattack stages};

\node at (5.5,1-0.65) {$\vdots$};
\node at (5.5,1.85) {$\vdots$};

\node at (9.5,1-0.65) {$\vdots$};
\node at (9.5,1.85) {$\vdots$};

\node at ($(k1.west)+(0.27,0.3)$) {$\checkmark$};
\node at ($(k2.west)+(0.27,0.3)$) {$\checkmark$};
\node at ($(k3.west)+(0.27,0.3)$) {$\checkmark$};


\draw[decorate, decoration={brace, amplitude=12pt, aspect=0.32}, thick] 
    (3,-1.5) -- (3, 3.25); 


\draw[dashed] (-2.2,-3.75) -- (-2.2,4);
\draw[dashed] (2.5,-3.75) -- (2.5,4);
\draw[dashed] (8,-3.75) -- (8,4);

\node(kc1) at (9.5,1-1.5) {  [1 0 0 0 0 0 0 0 0] };
\node(kc2) at (9.5,1) { [0 0 0 0 0 0 1 0 1]};
\node(kc3) at (9.5,2.5) {[0 0 0 0 0 1 0 0 0]};
\draw[dashed] (8.2,1-2.5) -- (11,1-2.5);
\node (sum) at (8.5,1-2.25) {$+$};

\node[below=3.75cm of kc3] (jobpie) {
    \piechart[0.5]{
    9/KA6, 
    10/KA9,
    4/KA1, 
    14/KA3, 
    6/KA4, 
    9/KA5, 
    36/KA7, 
    14/KA8}{}};

\end{tikzpicture}
}
\caption{Computation of the Knowledge Area distribution for individual job roles within the NICE framework. The figure illustrates the hierarchical mapping from job categories to job roles and a representative subset of Knowledge Descriptions associated with the \textit{Vulnerability Analysis} role. These KDs are labeled with KAs using ClassifyLM and aggregated to derive the corresponding KA distribution.}
\label{fig:nice_framework}
\end{figure}

In this work, we adopt the Knowledge Areas as a common frame of reference between cybersecurity curricula and workforce competencies. While PreprocessLM and ClassifyLM are used to map courses to KAs, we build on the methodology described by \cite{ramezanian2024cybersecurity} to compute the relative importance of the KAs for the overall job market, work role categories, and specific work roles. The overall procedure is visualized for the `Vulnerability Analysis' job role in \figref{nice_framework}, but the results for all NICE job roles can be found in \ref{app:KA_job_roles}. 

The NICE cybersecurity workforce framework \cite{NICE2025} is utilized to derive role-based weights for either individual job roles, job categories, or the overall cybersecurity labor market. As shown in \figref{nice_framework}, the framework is organized into five overarching work role categories.
Each of the 41 job roles is linked to exactly one category. 

Each job role in the NICE framework is characterized by three types of competency descriptors: Knowledge Descriptions, Skill Descriptions, and Task Descriptions. This study focuses exclusively on KDs, as these are directly relevant for curriculum analysis. KDs are standardized descriptions of knowledge, e.g. ``Knowledge of penetration testing tools and techniques", that are often applicable to multiple job roles.  

Most KDs were manually labeled with corresponding KAs by \citep{ramezanian2024cybersecurity}. However, an updated version of the NICE framework released in 2025 introduces significantly reformulated KDs, rendering many of the previous annotations obsolete. To ensure our role-based weightings accurately reflect the current standard, we adopt the most recent version of the NICE framework and automatically label all newly defined KDs using ClassifyLM. 

This comprehensive mapping allows us to compute a target KA distribution for any given job role. The calculation involves aggregating the KA distribution vectors of all KDs associated with a specific role and normalizing the resulting vector such that it sums to 100\%. \figref{nice_framework} provides a visual overview of this process and presents the final normalized KA distribution as a pie chart.
Note that it is also possible to compute the desired KA weighting for a job category by averaging the KA distributions of all roles contained in this category (see Section \ref{sec:job_to_KA}). Similarly, the aggregate KA distribution for the overall cybersecurity professional can be estimated by computing the mean KA distribution across all job categories, weighted by their respective labor market demand (see Section \ref{sec:KA_to_job}). 

\section{Validation Results } \label{sec:results}

This section presents a performance analysis of \CLLM{}. A quantitative comparison between \CLLM{} and human annotators is detailed in \secref{quant_cllm_humans}. Subsequently, \secref{qual_analysis} provides a qualitative evaluation, encompassing both a single-course case study and an analysis of three real-world curricula. Finally, \secref{future_proof} aligns \CLLM{} to workforce needs by estimating the job market relevance of each Knowledge Area and the desired KA distribution for the NICE work role categories.


\begin{figure}
    \centering
    \begin{subfigure}[b]{0.48\textwidth}
        \centering
        \begin{tikzpicture}[
            >=latex,
            every node/.style={font=\sffamily\Large},
            lab/.style={fill=white, inner sep=1pt, font=\small},
            scale=0.7
        ]
        \def\R{4} 
        
        \foreach \i/\name\names in {0/\CLLM{}/\CLLM{},1/Control A/Control A,2/P/$X_3$,3/V/$X_2$,4/S/$X_1$} {
          \node (\name) at ({90-72*\i}:\R) {\names};
        }

        \node [font=\small, left=1 mm of V] {avg: 62};
        \node [font=\small, above=1 mm of S] {avg: 72};
        \node [font=\small, right=1 mm of \CLLM{}] {avg: 61};
        \node [font=\small, above=1 mm of Control A] {avg: 64};
        \node [font=\small, right=1 mm of P] {avg: 78};
        
        \draw (\CLLM{}) -- node[lab, pos=0.50] {55} (Control A);
        \draw (Control A) -- node[lab, pos=0.50] {77} (P);
        \draw (P) -- node[lab, pos=0.50] {75} (V);
        \draw (V) -- node[lab, pos=0.50] {71} (S);
        \draw (S) -- node[lab, pos=0.50] {68} (\CLLM{});
        
        \draw (\CLLM{}) -- node[lab, pos=0.55] {44} (V);
        \draw (\CLLM{}) -- node[lab, pos=0.45] {72} (P);
        \draw (S) -- node[lab, pos=0.50] {62} (Control A);
        \draw (S) -- node[lab, pos=0.55] {87} (P);
        \draw (Control A) -- node[lab, pos=0.45] {59} (V);
        \end{tikzpicture}
        \caption{
        Comparison of Expert Group X ($X_1$, $X_2$, and $X_3$), Control Group A, and the proposed \CLLM{} on course data. The reported values indicate the percentage of extracted topics that share at least one overlapping assigned label between annotators. Percentages rounded to nearest integer.
        }
        \label{fig:curriculum_analysisv2}
    \end{subfigure}
    \hfill
    \begin{subfigure}[b]{0.48\textwidth}
        \centering
        \begin{tikzpicture}[
            >=latex,
            every node/.style={font=\sffamily\Large},
            lab/.style={fill=white, inner sep=1pt, font=\small},
            scale=0.7
        ]
        \def\R{4} 
        
        \foreach \i/\name\names in {0/\CLLM{}/\CLLM{},1/Control A/Control A,2/P/$X_3$,3/V/$X_2$,4/S/$X_1$} {
          \node (\name) at ({90-72*\i}:\R) {\names};
        }

        \node [font=\small, left=1 mm of V] {avg: 0.32};
        \node [font=\small, above=1 mm of S] {avg: 0.41};
        \node [font=\small, right=1 mm of \CLLM{}] {avg: 0.31};
        \node [font=\small, above=1 mm of Control A] {avg: 0.25};
        \node [font=\small, right=1 mm of P] {avg: 0.39};

        \draw (\CLLM{}) -- node[lab, pos=0.50] {0.21} (Control A);
        \draw (Control A) -- node[lab, pos=0.50] {0.30} (P);
        \draw (P) -- node[lab, pos=0.50] {0.38} (V);
        \draw (V) -- node[lab, pos=0.50] {0.42} (S);
        \draw (S) -- node[lab, pos=0.50] {0.43} (\CLLM{});
        
        \draw (\CLLM{}) -- node[lab, pos=0.55] {0.27} (V);
        \draw (\CLLM{}) -- node[lab, pos=0.45] {0.34} (P);
        \draw (S) -- node[lab, pos=0.50] {0.25} (Control A);
        \draw (S) -- node[lab, pos=0.55] {0.53} (P);
        \draw (Control A) -- node[lab, pos=0.45] {0.22} (V);
        \end{tikzpicture}
        \caption{Comparison of Expert Group X ($X_1$, $X_2$, and $X_3$), Control Group A, and the proposed \CLLM{} on course data. The values represent the inter-rater agreement in terms of Cohen's kappa. This value is between [-1,1], where higher is better.}
        \label{fig:curriculum_analysis}
    \end{subfigure}
    \caption{Comparison of Expert Group X ($X_1$, $X_2$, and $X_3$), the Control Group (Control A), and the proposed \CLLM{} model, evaluated on three randomly selected courses from each of the curricula of KTH, NTU, and CMU. Values for Control Group A represent the average over three annotators.}
    \label{fig:combined_analysis}
\end{figure}
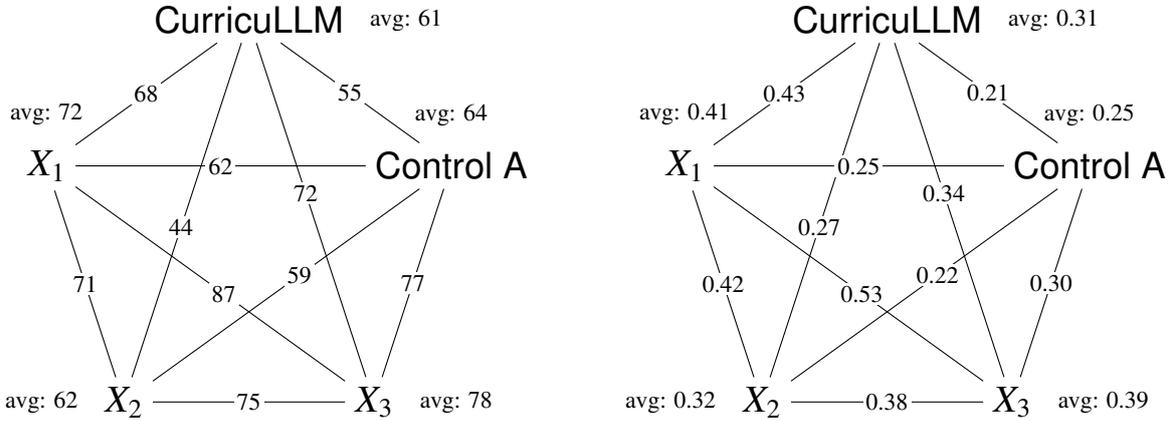

\subsection{Quantitative comparison: \CLLM{} vs Human Annotators}
\label{sec:quant_cllm_humans}


Since \CLLM{} was developed to automate  manual curriculum design, it is important to compare its performance to human experts. 
To this end, we conducted experiments involving three distinct groups of domain specialists, differing in their familiarity with the CSEC2017 framework and the type of data they annotated.

\begin{itemize}
    \item \textbf{Expert Group X} comprises  three domain experts, denoted as $X_1$, $X_2$, and $X_3$, who were familiar with the CSEC2017 framework prior to the study. These experts annotated \textit{all} data analyzed in this section.
    \item \textbf{Control Group A} consisted of 35 cybersecurity experts inquired to label real-world \textit{course data} during the Seminar of Swedish IT Security (SWITS) 2025 in Stockholm, Sweden. Each participant labeled up to seven extracted course topics. 
    \item \textbf{Control Group B} consisted of 34 experts consulted during the Secure Systems Demo Day 2025 in Espoo, Finland. Each participant in this group labeled 3 \textit{KDs} as defined in the NIST 2025 framework. 
\end{itemize}

The participants in Control Groups A \& B were provided with concise definitions of each Knowledge Area (as shown in \tabref{KAs}) and were asked to assign one or more KAs to a small subset of either real-world curriculum data or KDs.  
Section \ref{sec:curr_results_quant} presents the results of curriculum analysis, using Expert Group X and Control Group A as reference points for comparison with \CLLM{}. Section \ref{sec:KD_results} reports the results of labeling NIST KDs, evaluated with Expert Group X and Control Group B as a reference.



\subsubsection{Curriculum analysis of KTH, NTU, CMU}
\label{sec:curr_results_quant}
The level of agreement between \CLLM{} and human experts in classifying course data was evaluated as follows. First, a sample of nine courses was selected comprising three courses randomly chosen from each of the KTH, NTU, and CMU curricula. Next, PreprocessLM extracted 79 unique topics from these course descriptions, which served as input for both the ClassifyLM model and human annotators. 


The aggregated results of this comparative analysis are depicted in \figref{combined_analysis}. The raw data, i.e. the annotations of all 79 extracted topics by Expert Group X, Control Group A, and \CLLM{}, are listed in \tabref{network_systems_security} and \ref{app:extended_curriculum_analysis}. Here, \figref{curriculum_analysisv2} illustrates the percentage-wise agreement between all pair-wise combinations of annotators, where agreement was defined as the assignment of at least one common KA label to a given topic. For example, if one annotator assigned [1,4] and another assigned [2,4] to the same topic, the pair was considered to be in agreement.  

In addition, \figref{curriculum_analysis} reports the corresponding Cohen's kappa, which quantifies pair-wise agreement beyond chance. Cohen's kappa ranges from -1 to 1, where 0 indicates agreement equivalent to chance, and 1 signifies perfect agreement. This metric is stricter than percentage-wise agreement, as it requires complete consensus for a score of 1. 
The inter-annotator agreement was relatively high among Expert Group X, with the percentage-wise agreement ranging from 71--87\%, and Cohen's kappa values between 0.38 -- 0.53. As illustrated in \figref{curriculum_analysisv2}, it appears that the average agreement of \CLLM{} with Expert Group X (61\%) was comparable to agreement of Control Group A and Expert Group X (66\%). However, the results in \figref{curriculum_analysis} reveal a different trend: the mean Cohen's kappa compared to Expert Group X was significantly higher for \CLLM{} ($\kappa$ = 0.35) than for Control Group A ($\kappa$ = 0.26). This discrepancy indicates that, although participants in Control Group A tended to assign a greater number of KA labels per topic (increasing the chance of overlap), their classifications were less aligned with the CSEC2017 curriculum than those produced by \CLLM{}. This finding suggests that the KA labeling task requires familiarity with the CSEC2017 curriculum: while Control Group A participants were provided with basic definitions of each KA, this information alone was insufficient to achieve expert-level consistency with members of Expert Group X. 

Furthermore, the Cohen's kappa results indicate that \CLLM{} could automate course labeling with only a minor loss in performance relative to human experts. The mean agreement between \CLLM{} and Expert Group X ($\kappa$ = 0.35) is comparable to the agreement observed between Expert Group X annotators $X_2$ and $X_3$ ($\kappa$ = 0.38). In summary, \CLLM{} substantially outperformed Control Group A, consisting of cybersecurity experts previously unaware of the CSEC2017 framework, with only a marginal loss of performance compared to the CSEC2017-aware experts in Expert Group X ($X_1$, $X_2$, and $X_3$).


\subsubsection{NIST KD label analysis}
\label{sec:KD_results}

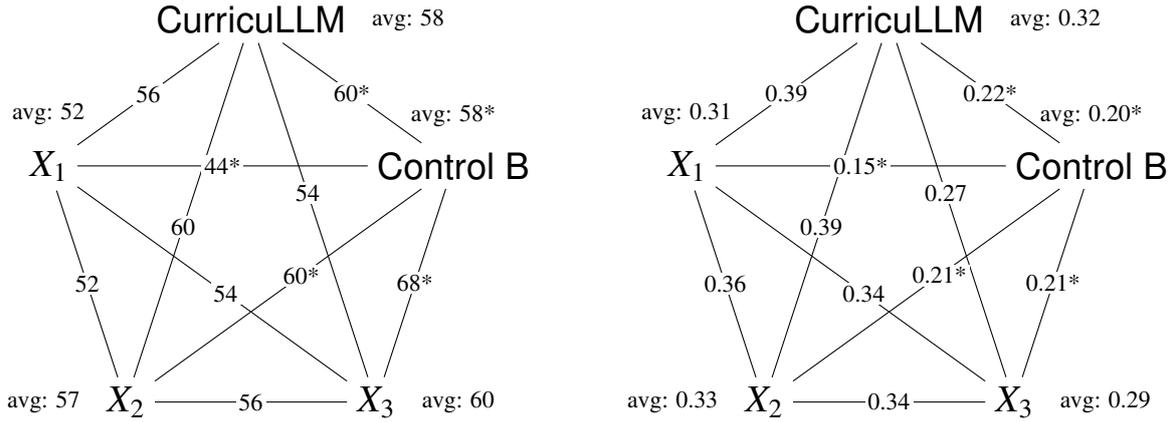
\begin{figure}[htbp]
    \centering
    \begin{subfigure}[b]{0.48\textwidth}
        \centering
        \begin{tikzpicture}[
            >=latex,
            every node/.style={font=\sffamily\Large},
            lab/.style={fill=white, inner sep=1pt, font=\small},
            scale=0.7
        ]
        \def\R{4} 
        
        \foreach \i/\name\names in {0/\CLLM{}/\CLLM{},1/Control B/Control B,2/P/$X_3$,3/V/$X_2$,4/S/$X_1$} {
          \node (\name) at ({90-72*\i}:\R) {\names};
        }
    
        \node [font=\small, left=1 mm of V] {avg: 57};
        \node [font=\small, above=1 mm of S] {avg: 52};
        \node [font=\small, right=1 mm of \CLLM{}] {avg: 58};
        \node [font=\small, above=1 mm of Control B] {avg: 58*};
        \node [font=\small, right=1 mm of P] {avg: 60};
        
        \draw (\CLLM{}) -- node[lab, pos=0.50] {60*} (Control B);
        \draw (Control B) -- node[lab, pos=0.50] {68*} (P);
        \draw (P) -- node[lab, pos=0.50] {56} (V);
        \draw (V) -- node[lab, pos=0.50] {52} (S);
        \draw (S) -- node[lab, pos=0.50] {56} (\CLLM{});
        
        \draw (\CLLM{}) -- node[lab, pos=0.55] {60} (V);
        \draw (\CLLM{}) -- node[lab, pos=0.45] {54} (P);
        \draw (S) -- node[lab, pos=0.50] {44*} (Control B);
        \draw (S) -- node[lab, pos=0.55] {54} (P);
        \draw (Control B) -- node[lab, pos=0.45] {60*} (V);
        \end{tikzpicture}
        \caption{Comparison of Expert Group X ($X_1$, $X_2$, and $X_3$), Control Group B, and the proposed \CLLM{} on KD data. The reported values indicate the percentage of extracted topics that share at least one overlapping assigned label between annotators. Percentages rounded to nearest integer. 
        }
        \label{fig:comparison_experts_swits_gpt_4o_mini}
    \end{subfigure}
    \hfill
    \begin{subfigure}[b]{0.48\textwidth}
        \centering
        \begin{tikzpicture}[
            >=latex,
            every node/.style={font=\sffamily\Large},
            lab/.style={fill=white, inner sep=1pt, font=\small},
            scale=0.7
        ]
        \def\R{4} 
        
        \foreach \i/\name\names in {0/\CLLM{}/\CLLM{},1/Control B/Control B,2/P/$X_3$,3/V/$X_2$,4/S/$X_1$} {
          \node (\name) at ({90-72*\i}:\R) {\names};
        }

        \node [font=\small, left=1 mm of V] {avg: 0.33};
        \node [font=\small, above=1 mm of S] {avg: 0.31};
        \node [font=\small, right=1 mm of \CLLM{}] {avg: 0.32};
        \node [font=\small, above=1 mm of Control B] {avg: 0.20*};
        \node [font=\small, right=1 mm of P] {avg: 0.29};

        \draw (\CLLM{}) -- node[lab, pos=0.50] {0.22*} (Control B);
        \draw (Control B) -- node[lab, pos=0.50] {0.21*} (P);
        \draw (P) -- node[lab, pos=0.50] {0.34} (V);
        \draw (V) -- node[lab, pos=0.50] {0.36} (S);
        \draw (S) -- node[lab, pos=0.50] {0.39} (\CLLM{});
        
        \draw (\CLLM{}) -- node[lab, pos=0.55] {0.39} (V);
        \draw (\CLLM{}) -- node[lab, pos=0.45] {0.27} (P);
        \draw (S) -- node[lab, pos=0.50] {0.15*} (Control B);
        \draw (S) -- node[lab, pos=0.55] {0.34} (P);
        \draw (Control B) -- node[lab, pos=0.45] {0.21*} (V);
        \end{tikzpicture}
        \caption{Comparison of Expert Group X ($X_1$, $X_2$, and $X_3$), Control Group B, and the proposed \CLLM{} on KD data. The values represent the inter-rater agreement in terms of Cohen's kappa. This value is between [-1,1], where higher is better.}
        \label{fig:kd_analysis_cohens}
    \end{subfigure}
    \caption{Comparison of Expert Group X ($X_1$, $X_2$, and $X_3$), the Control Group (Control B), and the proposed \CLLM{} model, evaluated on 50 Knowledge Descriptions extracted from the NICE 2025 framework. Values for Control Group B represent the average over five annotators, for a subset of 15 KDs (indicated with $*$).}
    \label{fig:combined_kd_analysis}
\end{figure}

To provide up-to-date workforce requirements and to assess the generalization capabilities of \CLLM{} beyond academic curricula, we conducted a second experiment focusing on the Knowledge Descriptions defined in the updated 2025 NICE Framework. 
This dataset differs fundamentally from course data, as each KD represents a specific cybersecurity skill or concept associated with \textit{professional} competency requirements. 
As discussed in \secref{role-basedweight}, the KDs form the foundation for computing the required KA distribution from the perspective of the cybersecurity job market, both for individual job roles and for the overarching workforce. 

To evaluate the performance of \CLLM{} on KD data, we analyzed 50 randomly selected KDs from the updated 2025 NICE Framework, ensuring that none overlapped with the entries used by \cite{ramezanian2024cybersecurity}. Each KD was preprocessed by removing the non-discriminative prefix `knowledge of' before classification. Subsequently, ClassifyLM was used to map each KD to one or more corresponding Knowledge Areas.


The results are summarized in \figref{combined_kd_analysis}. \figref{comparison_experts_swits_gpt_4o_mini} presents the percentage of overlapping KA assignments across annotator pairs, representing the proportion of KDs that shared at least one common KA label. \figref{curriculum_analysis} reports the inter-rater agreement measured using Cohen's kappa ($\kappa$), which accounts for agreement by chance. Note that a subset of 15 KDs was selected for comparative analysis with Control Group B (indicated as $*$ in \figref{combined_kd_analysis}).
Overall, both agreement metrics were considerably lower for KD data than for course data, with the mean percentage-wise agreement decreasing from 67\% to 56\%, and the mean Cohen's kappa dropping from 0.34 to 0.29. 
This outcome is expected, as the CSEC2017 framework was primarily designed to label courses, whereas KDs represent \textit{occupational} requirements. Nonetheless, while the absolute values declined, the relative trends remain consistent across datasets. 

Among Expert Group X, the mean percentage-wise agreement was 54\% and the mean Cohen’s kappa was $\kappa = 0.35$. The average agreement between \CLLM{} and Expert Group X was 57\% and $\kappa = 0.35$, effectively matching the mean inter-annotator reliability among the human experts. In contrast, the agreement between Control Group B and Expert Group X was substantially lower ($\kappa = 0.20$).
This pattern mirrors the findings from \secref{curr_results_quant}, where \CLLM{} also outperformed human control participants unfamiliar with the CSEC2017 framework. 

Taken together, these results demonstrate that \CLLM{} maintains expert-level consistency even when applied to non-academic, workforce-oriented data.

\subsection{Curriculum Analysis} \label{sec:qual_analysis}


To demonstrate the practical utility of \CLLM{} for curriculum design and analysis, we present a qualitative evaluation. This evaluation illustrates \CLLM{}'s functionality at the granular level of a single course (\secref{single_course}) and provides a comparative assessment of entire curricula from three institutions (\secref{whole_curriculum}).

\subsubsection{Single Course analysis: Building Networked Systems Security}
\label{sec:single_course}

\begin{table}[ht!]
\centering
\caption{Topics extracted from the Building Networked Systems Security Course (KTH), annotated with KAs by Control Group A, Expert Group X, and \CLLM{}}
\label{tab:network_systems_security}
\resizebox{.98\textwidth}{!}{
\begin{tabular}{m{6cm}ccccccc}
\toprule
  & \multicolumn{3}{c}{\textbf{Control Group A}} & \multicolumn{3}{c}{\textbf{Expert Group X}} & \textbf{\CLLM{}} \\
\cmidrule(l{1mm}r{1mm}){2-4}
\cmidrule(l{1mm}r{1mm}){5-7}
\textbf{Topics} & \textbf{\#1} & \textbf{\#2} & \textbf{\#3} & $X_1$ & $X_2$ & $X_3$ &  \\ \midrule
Building Networked Systems Security & 3--5,7 & 1,2,5,6 & 1,4,5 & 4 & 5 & 0,4 & 7 \\
\graybg Handling contemporary security problems for networked systems & 0,1,5--8 &  1,3-5 & 0,4--6 & 4 & 5 & 0,4 & 4 \\
Problem-based learning & 0,1,7 & 3,4,6 & -- & 0 & 0 & 0 & 0\\
\graybg Teamwork in cybersecurity & 0,1,4,6--8 & 0,3,6 & 1,5,7,8 & 7 & 6,7 & 0 & 7\\
Investigating requirements for networked systems security & 1,4 & 4,7 & 4 & 4 & 5 & 0--8 & 5 \\
\graybg Designing specifications for cybersecurity & 1,2,5,7 & 0,1,8 & 5 & 0,8 & 5 & 0,4--8 & 7 \\
Preparing solutions with professional tools & 0,2,7 & 5--8 & 1--3,5 & 0,7 & 2--4 & 0,3 & 7 \\
\graybg Critically assessing the efficiency of alternative solutions & 5-8 & 0 & 1--3 & 0 & 5 & 0--8 & 7 \\
\bottomrule
\end{tabular}
}
\end{table}

\begin{figure}[htb!]
\resizebox{.98\textwidth}{!}{
\begin{tikzpicture}[font=\scriptsize,node distance=3 cm]
\newcommand{\pieradius}{1.45}

\node (SWITS_PIE) {};
\node[right=of SWITS_PIE] (S_PIE) {};
\node[right=of S_PIE] (V_PIE) {};
\node[right=of V_PIE] (P_PIE) {};
\node[right=of P_PIE] (CLLM_PIE) {};

\node[align=center, font=\bfseries, above=1.5 cm of V_PIE] {};

\pie[
    pos=SWITS_PIE.center,
    radius=\pieradius,
    color={KA6,KA9,KA1,KA2,KA3,KA4,KA5,KA7, KA8},
    sum=auto
]{
    11/,
    11/,
    4/,
    7/,
    18/,
    16/,
    9/, 
    14/,
    9/
}

\pie[
    pos=S_PIE.center,
    radius=\pieradius,
    color={KA6,KA3,KA7, KA8},
    sum=auto
]{
   40/,
    30/,
    20/,
    10/
} 

\pie[
    pos=V_PIE.center,
    radius=\pieradius,
    color={KA6,KA1,KA2,KA3,KA4,KA5,KA7},
    sum=auto
]{
    9/,
    9/,
    9/,
    9/,
    45/,
    9/, 
    9/
}

\pie[
    pos=P_PIE.center,
    radius=\pieradius,
    color={KA6,KA9,KA1,KA2,KA3,KA4,KA5,KA7, KA8},
    sum=auto
]{
    25/,
    6/,
    6/,
    9/,
    16/,
    9/,
    9/, 
    9/,
    9/
}

\pie[
    pos=CLLM_PIE.center,
    radius=\pieradius,
    color={KA6, KA3,KA4,KA7},
    sum=auto
]{ 12/,
    12/,
    12/,
    62/
};

\node[align=left,font=\footnotesize,right=2 cm of CLLM_PIE] {
      \textbf{Security KA:} \\
   $\color{KA6}\bullet$ Miscellaneous\\
   $\color{KA9}\bullet$ Data\\
    $\color{KA1}\bullet$ Software\\
    $\color{KA2}\bullet$ Component\\
    $\color{KA3}\bullet$ Connection\\
    $\color{KA4}\bullet$ System\\
    $\color{KA5}\bullet$ Human\\
    $\color{KA7}\bullet$ Organizational\\
    $\color{KA8}\bullet$ Societal
    };

\node[below=1.5 cm of SWITS_PIE] (SWITS) {Control A};
\node[below=1.5 cm of S_PIE] (S) {Expert $X_1$};
\node[below=1.5 cm of V_PIE] (V) {Expert $X_2$};
\node[below=1.5 cm of P_PIE] (P) {Expert $X_3$};
\node[below=1.5 cm of CLLM_PIE] (CLLM) {\CLLM{}};

\end{tikzpicture}
}
\caption{The cumulative distribution of Knowledge Areas for the Building Network Security course offered at KTH, as annotated by Control Group A, Expert Group X ($X_1$, $X_2$, and $X_3$), and the proposed \CLLM{} framework. Percentages rounded to nearest integer.}
\label{fig:Network_Security_course}
\end{figure}
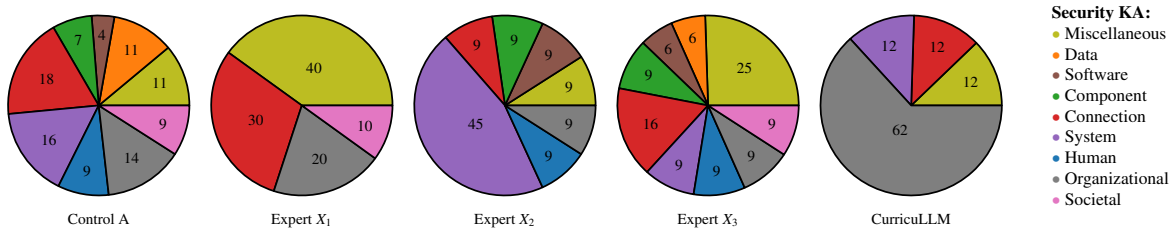

\CLLM{} was designed to analyze individual courses in a modular fashion, such that an overall curriculum can be built by selecting the courses that approximate the KA distribution required by the workforce. To illustrate this process, we examine a single course case study: the Building Networked Systems Security course offered at KTH (c.f. the case example presented in \secref{example}). Note that similar course-level data is available for 8 other courses, as outlined in \ref{app:extended_curriculum_analysis}. 

Our analysis focuses on publicly available course descriptions, as this information is available for most cybersecurity programs, ensuring \CLLM{} can be applied widely. Course descriptions, however, vary significantly in style. In some countries, such as Sweden, they are legally binding documents, whereas in other countries, course descriptions primarily serve as promotional material aimed at recruiting prospective students. Therefore, to be able to be processed further by a LLM, the course descriptions need to be standardized. As explained in \secref{datasets}, the descriptions are passed through \textit{PreprocessLM} to extract key pedagogical topics.

Following this procedure, the Building Networked Systems Security course description was processed by PreprocessLM, resulting in eight distinct pedagogical topics (listed in the leftmost column of \tabref{network_systems_security}). These topics were subsequently annotated by three  groups: (1) 35 participants from the SWITS conference (Control Group A), (2) our panel of curriculum-aware experts $X_1$, $X_2$, and $X_3$ (Expert Group X), and (3) the \CLLM{} model. 
In general, Control Group A exhibits little pair-wise overlap between experts \#1, \#2, and \#3. Consistent with the broader findings in \secref{curr_results_quant}, participants in Control Group A had a tendency to assign \textit{more} and \textit{less accurate} labels to each topic, resulting in a relatively high percentage-wise agreement with Expert Group X but lower Cohen's kappa scores. 

As displayed in \tabref{network_systems_security}, among Expert Group X, consensus was achieved only on the topic \textit{Problem-based learning}, which all assigned to the \textit{Miscellaneous} KA. Conversely, lower agreement was observed for more applied topics with weaker ties to cybersecurity core knowledge, such as `Critically assessing the efficiency of alternative solutions'. These discrepancies highlight the inherent subjectivity of the KA annotation process and suggest that some topics admit multiple valid interpretations. 


\CLLM{} showed perfect alignment with all experts on `Problem-based learning', agreement with Expert $X_1$ on `Handling contemporary security problems for networked systems', with Expert $X_2$ on `Investigating requirements for networked systems security' and partial alignment across multiple experts on `Teamwork in cybersecurity'. However, for topics more tangentially related to cybersecurity, such as `Preparing solutions with professional tools' expert diverged or opted for the Miscellaneous KA, while \CLLM{} assigned them to Organizational Security. This resulted in a larger representation of Organizational Security as compared to Expert $X_1$, $X_2$, and $X_3$. 
This discrepancy indicates that the experts utilize a different definition of Miscellaneous, i.e. `all competencies that are not directly cybersecurity related', while \CLLM{} is directly trained on the CSEC2017 description and the miscellaneous class.
For example, `Critically assessing the efficiency of alternative solutions' overlaps with the operational and tactical management Knowledge Unit contained in Organizational Security and little to no overlap with the Knowledge Units contained in Miscellaneous (see \tabref{KAs}). However, this topic could also be considered outside  the scope of cybersecurity and therefore be assigned to Miscellaneous, even though it does not relate to Knowledge Units contained in this KA.  

Then, moving towards curriculum design, the expert KA labels for the Building Networked Systems Security course are aggregated by taking the sum of all assigned labels to the topics contained in the course and normalizing the result to produce a probability distribution. The resulting normalized KA distributions are visualized as pie charts, as displayed in \figref{Network_Security_course}. As expected, the annotations from Control Group A approximate a uniform distribution over all KAs. Experts $X_1$ and $X_3$ assign the course the most to Miscellaneous and Connection Security, whereas expert $X_2$ favors System Security. Each of these annotations are defensible, since the course involves aspects of networks, systems, and also relates to softer skills that could be considered Miscellaneous. \CLLM{} assigns most topics to Organizational Security, and assigns the rest to Miscellaneous, Connection and System Security, in accordance with the experts in Expert Group X.

\subsubsection{Entire curriculum analysis: KTH, NTU and CMU}
\label{sec:whole_curriculum}

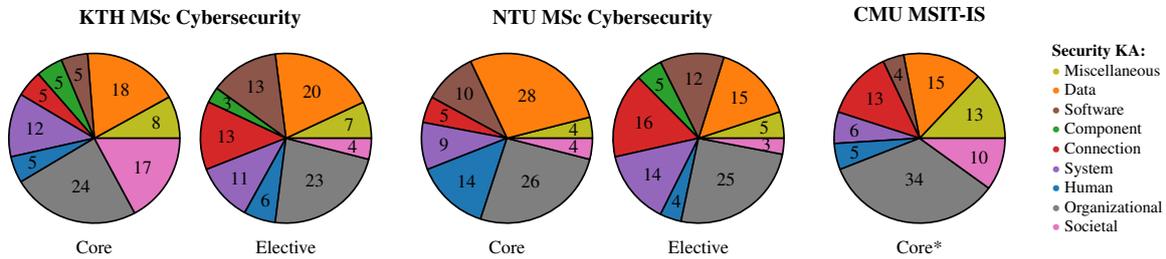
\begin{figure}[htb!]
\resizebox{.98\textwidth}{!}{
\begin{tikzpicture}[font=\small, node distance=3cm]
\newcommand{\pieradius}{1.45}

\node (KTH_CORE) {};
\node[right=of KTH_CORE] (KTH_ELECTIVE) {};
\node[right=3.5 cm of KTH_ELECTIVE] (NTU_CORE) {};
\node[right=of NTU_CORE] (NTU_ELECTIVE) {};
\node[right=3.5 cm of NTU_ELECTIVE] (CMU_CORE) {};

\pie[
    pos=KTH_CORE.center,
    radius=\pieradius,
    color={KA6,KA9,KA1,KA2,KA3,KA4,KA5,KA7,KA8},
    sum=auto
]{
8/, 
18/,
5/, 
5/, 
5/, 
12/, 
5/, 
24/, 
17/
}
\pie[
    pos=KTH_ELECTIVE.center,
    radius=\pieradius,
    color={KA6,KA9,KA1,KA2,KA3,KA4,KA5,KA7,KA8},
    sum=auto
]{
7/,
20/, 
13/, 
3/, 
13/, 
11/, 
6/, 
23/, 
4/
}

\pie[
    pos=NTU_CORE.center,
    radius=\pieradius,
    color={KA6,KA9,KA1,KA3,KA4,KA5,KA7,KA8},
    sum=auto
]{
4/,
28/, 
10/, 
5/, 
9/, 
14/, 
26/, 
4/
}

\pie[
    pos=NTU_ELECTIVE.center,
    radius=\pieradius,
    color={KA6,KA9,KA1,KA2,KA3,KA4,KA5,KA7,KA8},
    sum=auto
]{
5/,
15/, 
12/, 
5/, 
16/, 
14/, 
4/, 
25/, 
3/
}

\pie[
    pos=CMU_CORE.center,
    radius=\pieradius,
    color={KA6,KA9,KA1,KA3,KA4,KA5,KA7,KA8},
    sum=auto
]{
13/,
15/, 
4/, 
13/, 
6/, 
5/, 
34/, 
10/
}

\node[align=center, font=\bfseries, above=1.75 cm of $(KTH_CORE)!0.5!(KTH_ELECTIVE)$] {KTH MSc Cybersecurity};
\node[align=center, font=\bfseries, above=1.75 cm of $(NTU_CORE)!0.5!(NTU_ELECTIVE)$] {NTU MSc Cybersecurity};
\node[align=center, font=\bfseries, above=1.75 cm of CMU_CORE] {CMU MSIT-IS};

\node[below=1.5 cm of KTH_CORE] {Core};
\node[below=1.5 cm of KTH_ELECTIVE] {Elective};
\node[below=1.5 cm of NTU_CORE] {Core};
\node[below=1.5 cm of NTU_ELECTIVE] {Elective};
\node[below=1.5 cm of CMU_CORE] {Core*};

\node[align=left,font=\footnotesize,right=2 cm of CMU_CORE] {
      \textbf{Security KA:}\\
   $\color{KA6}\bullet$ Miscellaneous\\
   $\color{KA9}\bullet$ Data\\
    $\color{KA1}\bullet$ Software\\
    $\color{KA2}\bullet$ Component\\
    $\color{KA3}\bullet$ Connection\\
    $\color{KA4}\bullet$ System\\
    $\color{KA5}\bullet$ Human\\
    $\color{KA7}\bullet$ Organizational \\
    $\color{KA8}\bullet$ Societal
    };
\end{tikzpicture}
}
\caption{The cumulative percentage-wise distribution of Knowledge Areas as computed by \CLLM{}, for the core and elective part of the MSc programs in Cybersecurity and MSIT-IS offered at KTH, NTU and CMU, respectively. (*) For CMU, there was no fixed core program, so a representative sample core program was selected (more details in \secref{whole_curriculum}). }
\label{fig:entire_curriculum}
\end{figure}

\CLLM{} facilitates a fully automatic approach to curriculum analysis. Although \CLLM{} was initially conceived to analyze single courses and compose a curriculum based on these courses, it is equally capable of analyzing an entire curriculum by aggregating the KA labels of all courses contained in the curriculum. This is relevant to analyze to what extent a core curriculum covers job market demands, or whether provided conditionally elective courses are diverse enough for students to develop specialist competencies.  

To highlight the value of automated curriculum assessment, we analyzed three programs: the KTH MSc in Cybersecurity \citep{kth_ms_cybersecurity}, the NTU MSc in Cybersecurity \citep{ntu_msc_cybersecurity}, and the CMU MSc in Information Technology-Information Security (MSIT-IS) \citep{cmu_msit_is}. The results of this comparative analysis are presented in \figref{entire_curriculum}. For the KTH and NTU MSc, the results have been split into core and elective, where core indicates mandatory courses and elective indicates an aggregate of all conditionally (i.e. not free choice) elective courses, excluding degree projects. The CMU MSIT-IS degree's core consists mostly of conditionally elective courses. For the analysis provided here, a representative sample program fulfilling the degree requirements was selected.

A preliminary observation is that Organizational Security takes up a large proportion of all analyzed curricula. This is consistent with the single-course analysis provided in \secref{single_course}. Additionally, all three programs show a relatively light coverage of Component Security and Human Security, indicating a common curricular trend that prioritizes systemic and organizational aspects of cybersecurity over content related to hardware and human security.

KTH's core curriculum consists of 7 courses that together contribute to almost a third of the program: Theory and Methodology of Science, Theory of Science and Scientific methods in Cybersecurity, The Cybersecurity Engineer's Role in Society, Cybersecurity Overview, Cybersecurity in a Socio-Technical Context, Applied Cryptography, and Ethical Hacking. As depicted in \figref{entire_curriculum}, this results in a larger contribution of Organizational Security, Societal Security, Human Security and System Security in the program.
In contrast, the elective courses place greater emphasis on Connection Security and Software Security, with courses like `Building Networked Systems Security' and `Language-based Security'. 

NTU's core curriculum consists of 4 courses that contribute to 40\% of the curriculum; Computer Security, Application Security, Cryptography, Security and Risks Management. As compared to the core of KTH, this program has a stronger focus on Human Security and Data Security, while contributing less Societal Security. 
The 9 conditionally elective courses offered at NTU deliver a diverse palette of System Security courses (like Cyber physical systems security) and Connection Security (e.g. Network security). 

The analyzed CMU sample core program (consisting of Fundamentals of Telecommunications and Computer Networks, Browser Security, Introduction to Information Security, Information Security Risk Management, Information Security Policy and Management and Academic \& Professional Development) contributes to approximately 40 percent of the MSIT-IS program. Compared to the other programs, it shows less emphasis on technical areas like Data Security, System Security and Software Security, and more focus on Miscellaneous topics and Organizational Security.

\subsection{Results related to Job Market}
\label{sec:future_proof}

To assess how cybersecurity curricula can be aligned with current or emerging job market needs, we analyzed the correspondence between the KAs from the CSEC2017 framework and the cybersecurity work roles defined in the latest NICE 2025 framework \citep{NICE2025}. Specifically, this section examines (i) the overall relevance of each KA to the job market (\secref{KA_to_job}) and (ii) the desired KA distribution for each NICE job category (\secref{job_to_KA}). Together, these analyses provide insight in how curricula can be adapted to address the cybersecurity workforce gap. 

\subsubsection{Job Market Relevance of Knowledge Areas} \label{sec:KA_to_job}

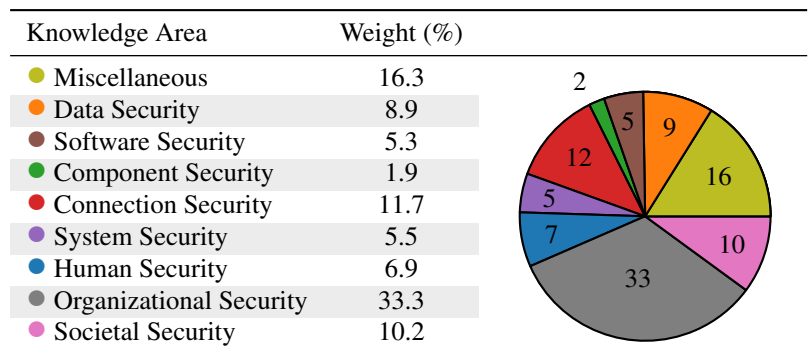
\begin{table}[ht!]
\caption{Relevance of KAs to the US job market based on demands between May 2023--April 2024 \citep{cyberseek}, expressed as a percentage (0--100\%). The KAs with larger percentages were more crucial to gain competency in the cybersecurity work roles with the highest demand in the aforementioned time period.}
\centering
\begin{tabular}{lcc}
\toprule
Knowledge Area  & Weight (\%) \\
\midrule
\bigbullet{KA6} Miscellaneous  & 16.3 & \multirow{9}*{
\begin{tikzpicture}
\pie[
    radius=1.65,
    color={KA6,KA9,KA1,KA2,KA3,KA4,KA5,KA7,KA8},
    sum=auto,
    ]
  { 16/, 9/, 5/, 2/, 12/, 5/, 7/, 33/, 10/ }
\pie[
    radius=1.65,
    color={KA6,KA9,KA1,KA2,KA3,KA4,KA5,KA7,KA8},
    sum=99, hide number]
  { 16/, 9/, 5/, 2/2 }
\pie[
    radius=1.65,
    color={KA6,KA9,KA1,KA2,KA3,KA4,KA5,KA7,KA8},
    sum=99]
  { 16/, 9/, 5/ }
\end{tikzpicture}
}\\
\graybg \bigbullet{KA9} Data Security & 8.9 \\
\bigbullet{KA1} Software Security   & 5.3\\
\graybg \bigbullet{KA2} Component Security & 1.9 \\
\bigbullet{KA3} Connection Security & 11.7\\
\graybg \bigbullet{KA4} System Security & 5.5 \\
\bigbullet{KA5} Human Security & 6.9\\
\graybg \bigbullet{KA7} Organizational Security  & 33.3 \\
\bigbullet{KA8} Societal Security  & 10.2 \\
\bottomrule
\end{tabular}
\label{tab:cybersecurity-weights}
\end{table}

To be able to design curricula targeted towards gaps in the cybersecurity workforce, it is important to know which aspects of the cybersecurity curriculum are most relevant for the job market. To this end, we extracted job demand data from the CyberSeek platform \citep{cyberseek}, which reports the number of job openings per NICE Cybersecurity Workforce job role in the US between May 2023 and April 2024. For each NICE job role, the demand was interpreted as a weight over the total number of listings. As explained in \secref{role-basedweight} and illustrated in \figref{nice_framework}, each NICE work role can be decomposed into Knowledge Descriptions, which are then mapped to Knowledge Areas contained in the CSEC2017 framework using the \CLLM{} framework. 
By propagating CyberSeek job demand weights through this mapping, we derived the aggregated relevance for each KA across all work roles, as depicted in \tabref{cybersecurity-weights}. It should be noted that the recently introduced Operational Technology Cybersecurity Engineering role, part of the Design \& Development category in NICE 2025, had no available job demand data at the time of this analysis.


Data in \tabref{cybersecurity-weights} reveal that Organizational Security forms the most dominant KA. This can in part be attributed to lexical patterns used in the revised KDs, e.g., the occurrence of words like `project' or `management', which tend to trigger classification under Organizational Security by \CLLM{}. This finding aligns with prior manual labeling efforts of KDs \citep{ramezanian2024cybersecurity}, which similarly identified Organizational Security as the most important KA. This consistency suggests that the trend reflects a genuine emphasis in the labor market rather than a model artifact.
Other high-weight KAs include Miscellaneous, Connection Security, Data Security, and Societal Security. The relatively large share of Miscellaneous is particularly noteworthy, since it indicates the need for cybersecurity experts with multidisciplinary knowledge. 

To clarify what conclusions can be drawn from \tabref{cybersecurity-weights}, we note that the data extracted from CyberSeek should be understood as describing market demand rather than the underlying normative needs of the cybersecurity ecosystem. Using the number of job listings as a measure of demand represents only one possible way to quantify workforce needs. Alternative options, such as weighting by the number of distinct organizations seeking a given role, could yield different demand distributions and highlight different KAs. 
Our analysis therefore solely estimates which Knowledge Areas best align with vacancies. From an information-theoretic perspective, an optimal match between an individual’s skill set and job market demands would therefore correspond to a probability distribution over all KAs that approximates the empirical distribution derived in \tabref{cybersecurity-weights}, for instance measured through the Kullback–Leibler divergence \citep{Cover-Thomas}.






\subsubsection{Desired KA distribution per NICE Job Category }
\label{sec:job_to_KA}

\begin{figure}[htb!]
\centering
\resizebox{.98\textwidth}{!}{
\begin{tikzpicture}[font=\small, node distance=3cm]
\newcommand{\pieradius}{1.45}

\node (OG_2017) {};
\node[right=of OG_2017] (DD_2017) {};
\node[right=of DD_2017] (IO_2017) {};
\node[right=of IO_2017] (PD_2017) {};
\node[right=of PD_2017] (IN_2017) {};

\node[below=5 cm of OG_2017] (OG_2025) {};
\node[right=of OG_2025] (DD_2025) {};
\node[right=of DD_2025] (IO_2025) {};
\node[right=of IO_2025] (PD_2025) {};
\node[right=of PD_2025] (IN_2025) {};

\node[align=center, font=\bfseries, above=1.5 cm of IO_2017] {NICE 2017};
\node[align=center, font=\bfseries, above=1.5 cm of IO_2025] {NICE 2025};

\pie[pos=OG_2017.center,
     radius=\pieradius,
     color={KA6,KA9,KA1,KA2,KA3,KA4,KA5,KA7,KA8},
     sum=auto]
    {23/, 2/, 4/, 9/, 13/, 6/, 1/, 32/, 10/}
\pie[pos=OG_2017.center,
     radius=\pieradius,
     color={KA6,KA9,KA1,KA2,KA3,KA4,KA5,KA7,KA8},
     sum=100.0,
     hide number]
    {23/, 2/, 4/, 9/, 13/, 6/, 1/1}
\pie[pos=OG_2017.center,
     radius=\pieradius,
     color={KA6,KA9,KA1,KA2,KA3,KA4,KA5,KA7,KA8},
     sum=100.0]
    {23/, 2/, 4/, 9/, 13/, 6/}
\pie[pos=OG_2017.center,
     radius=\pieradius,
     color={KA6,KA9,KA1,KA2,KA3,KA4,KA5,KA7,KA8},
     sum=100.0,
     hide number]
    {23/, 2/2}
\pie[pos=OG_2017.center,
     radius=\pieradius,
     color={KA6,KA9,KA1,KA2,KA3,KA4,KA5,KA7,KA8},
     sum=100.0]
    {23/}

\pie[pos=DD_2017.center,
     radius=\pieradius,
     color={KA6,KA9,KA1,KA2,KA3,KA4,KA5,KA7,KA8},
     sum=auto]
    {15/, 6/, 11/, 5/, 16/, 17/, 1/, 24/, 5/}
\pie[pos=DD_2017.center,
     radius=\pieradius,
     color={KA6,KA9,KA1,KA2,KA3,KA4,KA5,KA7,KA8},
     sum=100.0, hide number]
    {15/, 6/, 11/, 5/, 16/, 17/, 1/1}
\pie[pos=DD_2017.center,
     radius=\pieradius,
     color={KA6,KA9,KA1,KA2,KA3,KA4,KA5,KA7,KA8},
     sum=100.0]
    {15/, 6/, 11/, 5/, 16/, 17/}

\pie[pos=IO_2017.center,
     radius=\pieradius,
     color={KA6,KA9,KA1,KA2,KA3,KA4,KA5,KA7,KA8},
     sum=auto]
    {27/, 6/, 6/, 2/, 17/, 7/, 2/, 28/, 5/}
\pie[pos=IO_2017.center,
     radius=\pieradius,
     color={KA6,KA9,KA1,KA2,KA3,KA4,KA5,KA7,KA8},
     sum=100.0, hide number]
    {27/, 6/, 6/, 2/, 17/, 7/, 2/2}
\pie[pos=IO_2017.center,
     radius=\pieradius,
     color={KA6,KA9,KA1,KA2,KA3,KA4,KA5,KA7,KA8},
     sum=100.0]
    {27/, 6/, 6/, 2/, 17/, 7/}
\pie[pos=IO_2017.center,
     radius=\pieradius,
     color={KA6,KA9,KA1,KA2,KA3,KA4,KA5,KA7,KA8},
     sum=100.0, hide number]
    {27/, 6/, 6/, 2/2}
\pie[pos=IO_2017.center,
     radius=\pieradius,
     color={KA6,KA9,KA1,KA2,KA3,KA4,KA5,KA7,KA8},
     sum=100.0]
    {27/, 6/, 6/}


\pie[pos=PD_2017.center,
     radius=\pieradius,
     color={KA6,KA9,KA1,KA2,KA3,KA4,KA5,KA7,KA8},
     sum=auto]
    {12/ ,7/, 6/, 1/, 33/, 14/, 1/, 17/, 8/}
\pie[pos=PD_2017.center,
     radius=\pieradius,
     color={KA6,KA9,KA1,KA2,KA3,KA4,KA5,KA7,KA8},
     sum=99, hide number]
    {12/ ,7/, 6/, 1/, 33/, 14/, 1/1}
\pie[pos=PD_2017.center,
     radius=\pieradius,
     color={KA6,KA9,KA1,KA2,KA3,KA4,KA5,KA7,KA8},
     sum=99]
    {12/ ,7/, 6/, 1/, 33/, 14/}
\pie[pos=PD_2017.center,
     radius=\pieradius,
     color={KA6,KA9,KA1,KA2,KA3,KA4,KA5,KA7,KA8},
     sum=99, hide number]
    {12/ ,7/, 6/, 1/1}
\pie[pos=PD_2017.center,
     radius=\pieradius,
     color={KA6,KA9,KA1,KA2,KA3,KA4,KA5,KA7,KA8},
     sum=99]
    {12/ ,7/, 6/}


\pie[pos=IN_2017.center,
     radius=\pieradius,
     color={KA6,KA9,KA1,KA2,KA3,KA4,KA5,KA7,KA8},
     sum=100.0]
    {16/ ,31/, 5/, 2/, 12/, 6/, 0/, 16/, 12/}

\pie[pos=IN_2017.center,
     radius=\pieradius,
     color={KA6,KA9,KA1,KA2,KA3,KA4,KA5,KA7,KA8},
     sum=100.0, hide number]
    {16/ ,31/, 5/, 2/, 12/, 6/, 0/0}

\pie[pos=IN_2017.center,
     radius=\pieradius,
     color={KA6,KA9,KA1,KA2,KA3,KA4,KA5,KA7,KA8},
     sum=100.0]
    {16/ ,31/, 5/, 2/, 12/, 6/}

\pie[pos=IN_2017.center,
     radius=\pieradius,
     color={KA6,KA9,KA1,KA2,KA3,KA4,KA5,KA7,KA8},
     sum=100.0, hide number]
    {16/ ,31/, 5/, 2/2}

\pie[pos=IN_2017.center,
     radius=\pieradius,
     color={KA6,KA9,KA1,KA2,KA3,KA4,KA5,KA7,KA8},
     sum=100.0]
    {16/ ,31/, 5/}

\pie[pos=OG_2025.center,
     radius=\pieradius,
     color={KA6,KA9,KA1,KA2,KA3,KA4,KA5,KA7,KA8},
     sum=auto]
    {16/, 7/, 3/, 1/, 9/, 3/, 7/, 41/, 12/}
\pie[pos=OG_2025.center,
     radius=\pieradius,
     color={KA6,KA9,KA1,KA2,KA3,KA4,KA5,KA7,KA8},
     sum=99, hide number]
    {16/, 7/, 3/, 1/1}
\pie[pos=OG_2025.center,
     radius=\pieradius,
     color={KA6,KA9,KA1,KA2,KA3,KA4,KA5,KA7,KA8},
     sum=99]
    {16/, 7/, 3/}

\pie[pos=DD_2025.center,
     radius=\pieradius,
     color={KA6,KA9,KA1,KA2,KA3,KA4,KA5,KA7,KA8},
     sum=auto]
    {14/, 6/, 10/, 4/, 11/, 9/, 5/, 32/, 8/}

\pie[pos=IO_2025.center,
     radius=\pieradius,
     color={KA6,KA9,KA1,KA2,KA3,KA4,KA5,KA7,KA8},
     sum=auto]
    {20/, 11/, 4/, 2/, 12/, 5/, 8/, 29/, 10/}
\pie[pos=IO_2025.center,
     radius=\pieradius,
     color={KA6,KA9,KA1,KA2,KA3,KA4,KA5,KA7,KA8},
     sum=101, hide number]
    {20/, 11/, 4/, 2/2}
\pie[pos=IO_2025.center,
     radius=\pieradius,
     color={KA6,KA9,KA1,KA2,KA3,KA4,KA5,KA7,KA8},
     sum=101]
    {20/, 11/, 4/}

\pie[pos=PD_2025.center,
     radius=\pieradius,
     color={KA6,KA9,KA1,KA2,KA3,KA4,KA5,KA7,KA8},
     sum=auto]
    {15/, 11/, 3/, 1/, 16/, 5/, 6/, 33/, 9/}
\pie[pos=PD_2025.center,
     radius=\pieradius,
     color={KA6,KA9,KA1,KA2,KA3,KA4,KA5,KA7,KA8},
     sum=99, hide number]
    {15/, 11/, 3/, 1/1}
\pie[pos=PD_2025.center,
     radius=\pieradius,
     color={KA6,KA9,KA1,KA2,KA3,KA4,KA5,KA7,KA8},
     sum=99]
    {15/, 11/, 3/}

\pie[pos=IN_2025.center,
     radius=\pieradius,
     color={KA6,KA9,KA1,KA2,KA3,KA4,KA5,KA7,KA8},
     sum=auto]
   {12/, 21/, 3/, 2/, 9/, 7/, 5/, 28/, 12/}
\pie[pos=IN_2025.center,
     radius=\pieradius,
     color={KA6,KA9,KA1,KA2,KA3,KA4,KA5,KA7,KA8},
     sum=99, hide number]
   {12/, 21/, 3/, 2/2}
\pie[pos=IN_2025.center,
     radius=\pieradius,
     color={KA6,KA9,KA1,KA2,KA3,KA4,KA5,KA7,KA8},
     sum=99]
   {12/, 21/, 3/}

\node[below=1.5 cm of OG_2017, align=center] {Oversight \& \\ Governance (OG)};
\node[below=1.5 cm of DD_2017, align=center] {Design \& \\ Development (DD)};
\node[below=1.5 cm of IO_2017, align=center] {Implementation \& \\ Operation (IO)};
\node[below=1.5 cm of PD_2017, align=center] {Protection \& \\ Defense (PD)};
\node[below=1.5 cm of IN_2017, align=center] {Investigation (IN)};

\node[below=1.5 cm of OG_2025, align=center] {Oversight \& \\ Governance (OG)};
\node[below=1.5 cm of DD_2025, align=center] {Design \& \\ Development (DD)};
\node[below=1.5 cm of IO_2025, align=center] {Implementation \& \\ Operation (IO)};
\node[below=1.5 cm of PD_2025, align=center] {Protection \& \\ Defense (PD)};
\node[below=1.5 cm of IN_2025, align=center] {Investigation (IN)};

\node[align=left,font=\footnotesize, right=2 cm of $(IN_2017.center)!0.50!(IN_2025.center)$] {
      \textbf{Security KA:}\\
   $\color{KA6}\bullet$ Miscellaneous\\
   $\color{KA9}\bullet$ Data\\
    $\color{KA1}\bullet$ Software\\
    $\color{KA2}\bullet$ Component\\
    $\color{KA3}\bullet$ Connection\\
    $\color{KA4}\bullet$ System\\
    $\color{KA5}\bullet$ Human\\
    $\color{KA7}\bullet$ Organizational\\
    $\color{KA8}\bullet$ Societal
};

\end{tikzpicture}
}
\caption{The Knowledge Area distribution for each Job role category for the NICE 2017 \citep{NICE2017} (top) and NICE 2025 \citep{NICE2025} (bottom) frameworks. The numbers inside each pie chart represent the percentage-wise contribution of each KA (0--100\%), where KAs are color-coded as indicated in the legend on the right.}
\label{fig:job_role_KAs_combined}
\end{figure}
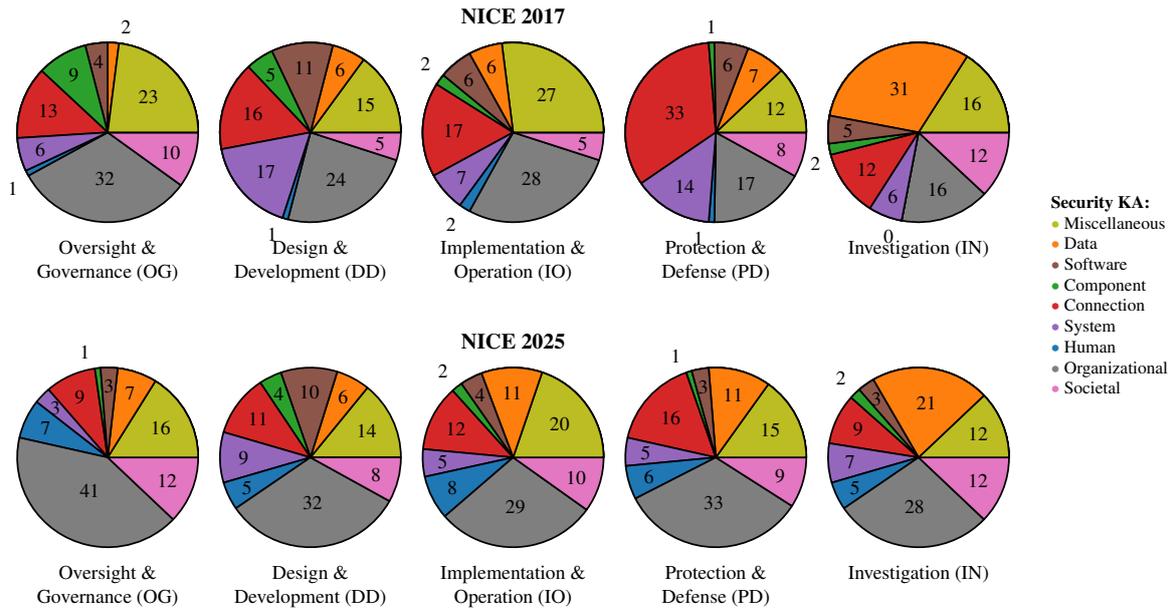

Beyond the aggregate market relevance of individual KAs, it is valuable to understand the specific KA profile required for different career paths. This information supports several use cases, such as identifying which elective courses best align with a student's desired career, or determining which NICE job categories a given curriculum prepares students most effectively for.

Therefore, we analyzed the desired KA distributions for every NICE job category using the methodology described in Section \ref{sec:role-basedweight}. The results for each NICE job role can be found in \ref{app:KA_job_roles}. To enable a longitudinal comparison of the requirements per NICE job category, we mapped the categories from NICE 2017 (Framework Components v1.0.0. \citep{NICE2017}) to their counterparts in NICE 2025 (Framework Components v2.0.0. \citep{NICE2025}). The largest difference between v1.0.0. and v2.0.0. was the removal of the `Collect \& Operate' and `Analyze' categories, which were taken up into the DoD Cyber Workforce Framework (DCWF) \citep{doddcwf}. Therefore, we discarded these categories from our analysis. 



Figure \ref{fig:job_role_KAs_combined} depicts the resulting KA distributions for the NICE job categories Oversight \& Governance (OG), Design \& Development (DD), Implementation \& Operation (IO), Protection \& Defense (PD), and Investigation (IN) for both 2017 (top) and 2025 (bottom). 
Several important distinctions between the 2017 and 2025 desired KA distribution can be noted. Firstly, Organizational Security is more prominent across all categories, reaching at least 28\% in each. At the same time, the contribution of Miscellaneous has shrunk significantly. As discussed in Section \ref{sec:qual_analysis}, this shift can be attributed to differences in labeling behaviour between human annotators and \CLLM{}: whereas human experts assigned topics they consider to be outside of the scope of cybersecurity to Miscellaneous, \CLLM{} tends to assign these to Organizational Security. 

Furthermore, Figure \ref{fig:job_role_KAs_combined} indicates that the distribution of KAs has become more uniform over time. In the NICE 2017 framework, the importance of KAs varied widely across workforce categories. For instance, Component Security contributed 9\% to OG, while it was considerably less prominent in other categories. In the 2025 Framework, the contribution of Component Security has equilibrated to a narrow range of 1\% and 4\% across all categories. 
Similarly, Connection Security took up between 12\% and 33\% in 2017 and between 9\% and 16\% in 2025. However, as can be observed in \ref{app:KA_job_roles}, the distributions for individual job roles still vary significantly. 

The largest changes are observable for the PD and IN workforce categories. Coincidentally, these categories have been modified most vigorously between 2017 and 2025. PD has absorbed several redefined work roles, including some formerly categorized under `Investigate'. At the same time, IN was reduced from three to two work roles. 
In contrast, IO, DD, and OG only changed marginally between 2017 and 2025. This indicates that the reported differences in KA distributions could in part be attributed to structural changes in the NICE framework. 
Additionally, the mean length of individual KDs has decreased from approximately 85 characters in 2017 to 52 characters in 2025, indicating a significant change to the phrasing of Knowledge Descriptions between 2017 and 2025. 

\section{Discussion} \label{sec:discussion}


Our results demonstrate the value of \CLLM{} as an end-to-end cybersecurity curriculum assessment tool. We contextualize these findings by outlining future research directions (\secref{future_directions}), 
and addressing this study's limitations (in \secref{limitations}).

\subsection{Future directions} \label{sec:future_directions}
Future work may extend the applicability and generality of our proposed framework. A natural next step is the release of \CLLM{} as a tool capable of extracting course listings, curriculum structures, and course descriptions from the websites of established cybersecurity programs. Such functionality would enable large-scale data acquisition and support comparative analysis across institutions. 

Another direction involves enhancing the framework's capacity to model a broader range of constraints on cybersecurity curricula, such as practical scheduling conflicts, and credit dependencies. Furthermore, \CLLM{} could be expanded to assess a student's prior knowledge based on the completed parts of their curriculum, such that only courses of the appropriate level and with content not previously covered in their program are recommended. 


Because \CLLM{} is not inherently constrained to the NICE workforce framework, it can be readily aligned with alternative workforce taxonomies, such as the European Cybersecurity Skills Framework \citep{enisa_skills_framework_2022}. More broadly, the pipeline could be adapted to other academic domains for which well-established curricular guidelines are available. For instance, the 2023 Computer Science Curricula  \citep{kumar2024computer} provides a suitable foundation for such an extension. 

Finally, the weighting scheme used to represent the relative importance of Knowledge Areas based on regional demands (as introduced in Section \ref{sec:KA_to_job}) could be recomputed based on other geographical locations, such as those documented for the Finnish cybersecurity sector \citep{lehto2022development}. Furthermore, projected job market demands can be taken into account to support the development of curricula that remain aligned with evolving professional requirements. 



\subsection{Limitations} \label{sec:limitations}
Although the \CLLM{} framework provides a structured approach for connecting curriculum content to workforce requirements, several limitations should be acknowledged. 
A central challenge in mapping curriculum content to Knowledge Areas is the inherently low \textit{inter-rater agreement} observed among human annotators. Because the KAs partially overlap and are not mutually exclusive by design, human judgments often diverge, making it difficult to establish a definitive ground truth. This complicates both model training and evaluation. Consequently, the objective of \CLLM{} was not to maximize conventional metrics, but to approximate the level of agreement typically achieved by human annotators. 


Another limitation arises from the fact that the current pipeline is exclusively based on \textit{course descriptions}. While these descriptions are widely accessible and offer a practical entry point for large-scale data collection, they provide only a partial representation of course content. Other sources of information, such as learning objectives or sample educational material, could yield a more comprehensive understanding of the scope of the course and its content. However, such data is often unavailable due to copyright restrictions or institutional access constraints. 
Moreover, our reliance on course descriptions may introduce bias, as these descriptions are often designed to appeal to prospective students, and may overemphasize the societal relevance of the course rather than the core technical competencies covered. 





\section{Conclusions} \label{sec:conclusion}
In this work, we introduced \CLLM{}, a novel LLM-based framework to automate the analysis and personalization of cybersecurity curricula. Our approach directly addresses the gap between academic programs and industry needs by enabling data-driven alignment between curricula and industry standards. 
The core of our method is a two-tier approach: PreprocessLM transforms input data into a format suitable for machine learning and ClassifyLM classifies the data into the nine Knowledge Areas, where eight were defined by the CSEC2017 framework and one defined by \cite{ramezanian2024cybersecurity}. Among several NLP methods evaluated, the BERT-based model demonstrated the best performance, providing a solid foundation for large-scale automated curriculum analysis. 

Our quantitative results show that \CLLM{} serves as a practical and effective tool for cybersecurity curriculum assessment. It accurately maps educational content to the CSEC2017 Knowledge Areas, achieving an inter-rater agreement comparable to human annotators, with minimal manual intervention or performance loss.
Furthermore, we show how \CLLM{} can be applied at different levels of granularity, from individual courses to entire programs, to automatically identify the distribution of KAs within a curriculum. 

An important aspect of \CLLM{} is its ability to evolve with the field, resulting in adaptive and workforce-aligned curricula. By integrating the newly released 2025 NICE Workforce Framework, \CLLM{} reflects the most recent definitions of cybersecurity competencies and work categories. This enables quantifying the job market relevance of individual Knowledge Areas based on current listings. By identifying which KAs should be emphasized in education according to current job market demands, we contribute to closing the cybersecurity workforce skills gap. 

The methodological foundation developed for \CLLM{}, including dataset preparation, model training, and validation on real-world curricula, offers a systematic approach to applying LLMs to curricula development. Potential future work includes incorporating regional labor market data, expanding towards other related disciplines, and anticipating emerging skill demands. 
By bridging the gap between curricula and the dynamic cybersecurity workforce landscape, this work contributes to building more automated,  industry-relevant, and personalized educational programs.

\section*{Acknowledgments}
Funding for this work was made available by Cyber Security Education Cooperation Network Project Finnish Ministry of Education and Culture (Grant Number: OKM/60/522/2022), Swedish Research Council (VR) under grant 2021-05621, and the Swedish Knowledge Foundation (KKS).

All computations were executed on a Tesla T4 GPU, provided by the National Academic Infrastructure for Supercomputing in Sweden (NAISS), partially funded by the Swedish Research Council through grant agreement no. 2022-06725.

\section*{Data Availability statement}
All datasets used in this work, as well as the Python scripts used to generate our synthetic CSEC2017 dataset, are publicly accessible on our GitHub repository: \url{https://anonymous.4open.science/r/LLMedu-227A}. 

\section*{Declaration of Interest statement}
Arthur Nijdam reports financial support and equipment were provided by Swedish Research Council. Harri Kähkönen and Valtteri Niemi report financial support was provided by Finnish Ministry of Education and Culture. Sara Ramezanian reports financial support was provided by Swedish Knowledge Foundation (KKS). Paul Stankovski Wagner declares not to have any known competing financial interests or personal relationships that could have appeared to influence the work reported in this paper.





\bibliographystyle{elsarticle-harv}
\bibliography{ref}

\appendix
\section{Fine-tuning Strategy} \label{app:finetuning}

In this section, our goal is to identify a suitable base model to serve as \textit{ClassifyLM} and fine-tune it for the task of KA classification using our curated dataset. We detail the baseline methods analyzed (\secref{baselines}), the evaluation metrics used for comparison (\secref{metrics}), and the results of our experimental analysis of the candidate models (\secref{model_selection}).

\subsection{Baselines} \label{sec:baselines}

We consider the following NLP methods for the ClassifyLM task: 
\begin{itemize}
    \item Zero-shot Learning: In zero-shot learning, we simply \textit{prompt} a LLM without domain-specific fine-tuning. We selected DeepSeek-V3 \citep{liu2024deepseek} and ChatGPT-4o-mini \citep{ChatGPT} for this baseline, since both models are widely available and relatively cost-effective. Additionally, we evaluate Qwen-2.5b-7B-Instruct \citep{qwen2.5} (the model adopted for PreprocessLM), to motivate that different models were used for PreprocessLM and ClassifyLM. The specific prompt used for zero-shot classification is detailed in Figure \ref{fig:llmprompt2} in \ref{app:zero_shot_classification}. 

    \item Random Forest (RF): RFs are a well-known traditional machine learning method. In contrast to deep learning methods that learn directly from raw text, RFs require feature extraction. Here, we use both Bag-of-Words (BoW) and Term-Frequency-Inverse Document Frequency (TF-IDF), that assign numerical values to words based on how frequently they appear in the training set. 

    \item LSTM: To establish a direct comparative baseline with the most relevant related work by Dzurenda et al. \citep{dzurenda2024enhancing}, we utilized the core architecture described in their paper. This model employs the standard BERT \citep{devlin2019bert} tokenizer for text processing, followed by a unidirectional Long Short-Term Memory (LSTM) network for the final classification. Note that it was not possible to use Dzurenda et al.'s trained model directly, as it was optimized for classifying 12 European Union Agency for Cybersecurity (ENISA) work profiles rather than the 9 CSEC2017 Knowledge Areas. Therefore, we adapted their architecture and trained it on our dataset with the recommended learning rate of 0.001 for 20 epochs. 

    \item Fine-tuned Transformer Models: We experimented with several transformer variants, namely BERT \citep{devlin2019bert}, ALBERT \citep{lan2019albert}, RoBERTa \citep{liu2019roberta}, and DistilBERT \citep{sanh2019distilbert}. BERT serves as a foundational benchmark. DistilBERT offers a more parameter-efficient alternative, reducing the model size by approximately 40\% through knowledge distillation. RoBERTa improves upon BERT with optimized training techniques like dynamic masking and extended training, while ALBERT introduces mechanisms to minimize the memory footprint. All models were fine-tuned for a maximum of 20 epochs using a sigmoid output activation and the Binary Cross Entropy loss to accommodate the multi-label classification setting. 
\end{itemize}


\subsection{Metrics} \label{sec:metrics}

The task of assigning course descriptions to one or more KAs is a multi-label, multi-class classification problem. 
While the accuracy provides an intuitive measure of overall prediction correctness, it is not well-suited for multi-label classification. It is highly affected by class imbalance: models can achieve a high accuracy solely by performing well on the KA with the highest number of samples. Accuracy also treats all errors equally and does not distinguish between false positives and false negatives, even though these could have different implications for curriculum analysis. 
Therefore, to enable a more nuanced performance evaluation, we do not use the accuracy, but instead compute the macro-averaged precision, recall, and F1 score. Here, `macro-averaged' indicates that metrics are computed per KA and then averaged, such that KAs with fewer samples contribute as much to the final score as KAs with more samples. The definitions of the metrics used are as follows: 

\begin{itemize}
    \item Precision:  
    The precision or \textit{true positive rate} measures the proportion of correctly predicted positive labels among all predicted positives for a given class. We report the \textit{macro-averaged precision}, which computes the metric independently for each KA and then takes the mean, assigning equal weight to all KAs. This is defined as:
    \begin{equation}
        Precision =  \frac{1}{C} \sum_{i=1}^{C} \frac{TP_i}{TP_i + FP_i}
    \end{equation}
    Here, $TP_i$ represents the number of True Positives (TPs) for KA $i$, while $FP_i$ stands for the number of False Positives (FPs). $C$ is the total number of classes, which is equal to 9 in the case of KAs. 
    Macro-averaged precision yields a value between 0 and 1, where a high value indicates a low rate of false positives across the model's predictions.
    \item Recall: The recall or \textit{sensitivity} quantifies the proportion of actual positives that have been correctly identified as positive by the model. Analogous to precision, we report the \textit{macro-averaged recall}:
    \begin{equation}
        Recall = \frac{1}{C} \sum_{i=1}^{C} \frac{TP_i}{TP_i + FN_i}
    \end{equation}
    Here, $FN_i$ stands for the number of False Negatives (FNs) for a given KA $i$.
    The macro-averaged recall yields a value between 0 and 1, where a high recall can be achieved when the model has few false negatives. 
    
    \item F1 score: The F1 score is the harmonic mean of the precision and the recall, addressing scenarios where one of these two metrics dominates. We report the \textit{macro-averaged F1 score}, calculated as the mean of the per-class F1 scores: 
    \begin{equation}
    \text{F1} =  \frac{1}{C} \sum_{i=1}^{C} \frac{2 \, TP_i}{2 \, TP_i + FP_i + FN_i}.
\end{equation}

    The F1 score ranges from 0 to 1, where a higher value indicates superior performance by jointly considering precision and recall. Although the F1 score is the harmonic mean of precision and recall, its macro-average is not. Consequently, the macro-averaged F1 score can be lower than the macro-averaged precision and recall.
\end{itemize}

\subsection{Model Selection} \label{sec:model_selection}
\begin{table*}[ht!]
    \small
    \caption{ 
    The classification performance of various NLP methods on the fine-tuning dataset. Column 1 gives an overview of the NLP model class used, Column 2 provides model specifications. The performance metrics used are the macro-averaged Precision (0--1), Recall (0--1), and F1 score (0--1), as defined in Section \ref{sec:metrics} The reported scores were computed using 10-folds cross validation and have been averaged over 5 random seeds. }
    \label{tab:Baseline_results}
    \centering
    \begin{tabular}{lcccc}
    \toprule
    Category & Specification  & Precision ($\uparrow$) & Recall ($\uparrow$) & F1 score ($\uparrow$) \\ 
\midrule
    Zero-shot & Qwen-2.5b-7B-Instruct \citep{qwen2.5}
    & 0.29 &  0.44 & 0.32\\ 
     & ChatGPT-4o-mini \citep{ChatGPT}
     & 0.40 & 0.33  &  0.29 \\ 
     & DeepSeek-V3 \citep{liu2024deepseek} 
     & 0.38& 0.49 & 0.41 \\ 
\graybg Random Forest & Bag-of-words
    & 0.67 & 0.44 & 0.53\\
\graybg & TF-IDF 
     & 0.71 & 0.44 & 0.54 \\
    LSTM  & BERT tokenizer \citep{dzurenda2024enhancing} 
    & 0.14 & 0.13 & 0.13 \\
\graybg Transformers
     & ALBERT \citep{lan2019albert} 
     & 0.68 & 0.48 & 0.55 \\
    
\graybg     & DistilBERT \citep{sanh2019distilbert} 
     & 0.71 & 0.57  & 0.62 \\
\graybg     & RoBERTa \citep{liu2019roberta}
     & 0.69 & 0.58 & 0.63 \\
\graybg   \hspace{.75cm}\textbf{(selected)$\longrightarrow$}   & \textbf{BERT \citep{devlin2019bert}} 
      & \textbf{ 0.71} & \textbf{ 0.59} & \textbf{0.64} \\
\bottomrule
    \end{tabular}
\end{table*}

\tabref{Baseline_results} gives an overview of the cross-validation scores of all baseline NLP methods mentioned in \secref{baselines}, as computed on the fine-tuning dataset (described in \secref{datasets}). 
To ensure a robust performance estimate, we use \textit{k-fold cross validation}. This technique partitions the fine-tuning dataset, consisting of synthetic CSEC2017 data and NIST Knowledge Descriptions as described in the previous subsection, into $k$ non-overlapping folds. In each of the $k$ iterations, the model is trained on $k-1$ folds (corresponding to 90\% of the data for $k=10$) and validated on the hold-out fold (the remaining 10\%). In this way, it is possible to compute the performance of the model on the entire fine-tuning dataset. To balance the bias and variance trade-off associated with this process, the recommended setting for $k$ is 10. We adopt $k=10$ and split the training dataset into 10 non-overlapping 90/10 train/val splits. 

We first evaluate zero-shot classification performance using Qwen-2.5b-7B-Instruct (which also serves as the backbone for PreprocessLM), and the state-of-the-art DeepSeek-V3 and ChatGPT-4o-mini models. An identical prompt, as indicated in \figref{llmprompt2} in \ref{app:zero_shot_classification}, was used for all models to ensure a fair comparison. 
As mentioned in earlier sections, we hypothesized that the tasks designed for PreprocessLM and ClassifyLM cannot be executed by the same model. From the data in \tabref{Baseline_results} we can see that the Qwen-2.5b-7B-Instruct model does not have a good off-the-shelf performance for KA classification, a limitation that cannot be remedied via fine-tuning due to the small size of our dataset. The more powerful zero-shot DeepSeek-V3 and ChatGPT-4o-mini models also achieve low scores, demonstrating the limitations of general-purpose LLMs without domain-specific fine-tuning.

Random Forest (RF) based methods showed moderate improvements over zero-shot LLM prompting, which suggests that the features extracted from the fine-tuning dataset possess predictive value. Nonetheless, the limited performance of the RF models indicates that these models struggle to capture the complex semantics inherent to cybersecurity concepts. 
The baseline method proposed by \cite{dzurenda2024enhancing}, based on a unidirectional LSTM trained on pre-tokenized text, performed significantly worse than all other approaches. 
However, this can in part be attributed to the fact that the hyperparameter settings and training methodology were not optimized for our dataset, but set to the default values recommended by  \cite{dzurenda2024enhancing}. A bidirectional LSTM or a deeper neural network architecture could have improved the performance of the LSTM-based approach, but these architectural changes are beyond the proposed methodology of Dzurenda et al. and therefore considered outside of the scope of this paper. 

In contrast, Transformer-based architectures substantially outperformed other approaches, with BERT achieving an F1 score of 0.64. We attribute this superior performance to the self-attention mechanism in transformer models, which allows them to weigh the importance of all words in a context simultaneously, capturing long-range contextual information more effectively than LSTMs or RFs. While transformer-based models generally outperformed traditional methods, BERT emerged as the best candidate. 
Based on these results, we therefore designate the fine-tuned BERT model as ClassifyLM. 

\section{Constrained Optimization for curriculum composition} \label{sec:constr_optimization}

After fine-tuning the ClassifyLM model as detailed in \secref{LL2_finetuning}, it can be used to estimate the KA distribution of any given course based on its course description. The model produces a probability distribution over the CSEC2017 Knowledge Areas, providing a representation of the course's cybersecurity coverage. 

As established in \secref{role-basedweight}, target requirements from the NICE framework for job roles, job categories, and the overall market, are similarly encoded as a probability vector over KAs. This shared vector representation enables the direct alignment of courses with cybersecurity workforce demands, framing the task of optimal course selection as a constrained optimization problem.


To design a curriculum aligned with job  market requirements, it is important to consider the structural constraints of the curriculum. In particular, factors such as the total number of credits in the program, and the desired distribution between core and elective credits effectively \textit{constrain} the optimization process. For example, the two-year MSc in Cybersecurity at KTH consists of 120 European Credit Transfer and Accumulation System (ECTS) credits, of which 39.5 ECTS correspond to mandatory core courses, 30 ECTS to conditionally elective courses, 30 ECTS are dedicated to the master thesis project and the remaining 20.5 ECTS are free choice electives. 
As presented in the case example (\secref{example}), \CLLM{} can be used to identify the optimal set of conditional electives for a specific target job role. The program offers 12 elective courses to first-year students, of which students are required to select at least 4. 
Although \CLLM{} can also be used to recommend free-choice electives, the present scenario focuses on optimizing the selection of conditional electives for simplicity. We formulate the optimization problem for the present scenario as a second-order constrained optimization problem, with the objective of aligning the student's composed curriculum as closely as possible with the desired KA distribution of the target job role. 

Here, we let
\begin{itemize}
    \item $m$: Number of KAs, $k$: Number of elective courses to be selected, $n$: Total number of available electives, $c_M$: Total number of mandatory credits. 
    \item $c_i$ the credit attributed to course $i$. 
    \item $E_{i,j} \in [0,1]$: Contribution of KA $j$ to elective course $i$. This is a normalized probability vector, such that $\sum_{j=1}^mE_{i,j}=1$. 
    \item $T_j \in [0,1]$: Target KA distribution of the desired job role. This is a normalized probability vector, subject to $\sum_{j=1}^mT_{j}=1$. 
    \item $x_i \in \{0,1\}$: Binary decision variable indicating whether elective $i$ is selected. 
    \item $\mathcal{M} \subset \mathbb{R}^m$: The aggregated KA vector of the mandatory courses. 
\end{itemize}
Such that the MIP formulation becomes 
\begin{align}
\text{minimize} \quad & \sum_{j=1}^{m} \left| 
\frac{c_M}{c_M+\sum_{i=1}^nc_ix_i}\mathcal{M}_j + \frac{1}{c_M+\sum_{i=1}^nc_ix_i}\sum_{i=1}^{n} c_ix_i E_{i,j} - T_j \right| \\
\text{subject to} \quad & \sum_{i=1}^{n} x_i = k \\
& x_i \in \{0, 1\} \quad \forall i \in \{1, \dots, n\}
\end{align}

Each course (mandatory or elective) is represented as a continuous $m-$dimensional probability vector indicating the proportional contribution of each KA to the course. Specifically, each elective $i \in {1, \dots, n}$ is associated with a normalized contribution vector $E_i = [E_{i,1}, \dots, E_{i,m}]^\top$. 
The mandatory curriculum component is similarly represented by a vector $\mathcal{M} \in [0,1]^m$ and each elective course $i \in \{1, ..., n\}$ has a KA contribution vector $E_i\subset \mathbb{R}^m$. The desired job profile is represented as a target KA distribution $T \subset \mathbb{R}^m$, obtained from Table 3 in 
\citep{ramezanian2024cybersecurity}. Binary variables $x_i\in \{0,1\}$ are introduced to indicate whether a given elective course $i$ has been selected. The total number of electives selected is constrained to $k$ using constraints (5) and (6). Additionally, because mandatory and elective courses carry different credit values at the program at KTH, we aggregate KA distributions in a credit-weighted manner. We let $c_M$ denote the total number of mandatory credits and $c_i$ the credits of elective $i$. 

The KA distribution of the student's entire curriculum is then computed as the sum of the mandatory component, $\mathcal{M}$, and the contributions of the selected electives: $\sum_{j=1}^{m}\sum_{i=1}^{n}c_ix_iE_i$, weighted by the number of credits contained in each. 
The objective (4) is then to minimize the absolute deviation between the KA profile of the student's curriculum and the target profile $T$. This ensures that the selected electives fill the knowledge gap left by the mandatory curriculum as closely as possible.

We next examine to what extent this optimization problem should be reformulated to accomodate alternative curriculum constraints. 
The one-year MSc in Cybersecurity at Nanyang Technological University (NTU) consists of 30 Academic Units (AU), of which 12 AUs are reserved for mandatory courses and 6 AUs are dedicated towards a capstone project. The remaining 12 AUs are conditionally elective, where students choose four elective courses from a pool of nine. Because of the great structural similarity of this program with the program offered at KTH, \CLLM{} can be applied to the NTU master in a similar fashion as mentioned before. 

The two-year M.S. in Information Technology - Information Security (MSIT-IS) offered at Carnegie Mellon University (CMU) consists of 75 core units, 36 free program electives, and 12 restricted electives. While the restricted electives can be chosen in a similar way as posed in the optimization formulation for KTH, the core units have a more complicated structure that we consider outside the scope of our current work. The core units are separated into a networking core (12 units, one course selected out of four options), an MSIT-IS core (12 units, one course selected from three options), a Security core (12 units, fixed), a Business and management component (12 units, comprising either two fixed courses or one alternative course), and a practical project worth 24 units. 
Additionally, we have not analyzed constraints such as the sequentiality of courses (e.g., Advanced calculus requires prior completion of Basic Calculus), practical constraints such as clashing timeslots, and the student's prior knowledge.

\section{Curriculum Analysis extended results}
\label{app:extended_curriculum_analysis}
For each of the 3 curricula analyzed in \secref{job_to_KA} (KTH, NTU, CMU), we randomly selected 3 courses to be analyzed in further detail. Here, the results for the Building Networked Systems Security (KTH) course are provided earlier in the paper in \tabref{networked_systems_security} in  \secref{KA_to_job}. 
The rest of the results can be found in the following tables:  
Ethical Hacking (KTH), \tabref{ethical_hacking}, Language-Based Security (KTH), \tabref{language_based_security}, Network Security (NTU), \tabref{network_security}, Software Security (NTU), \tabref{software_security}, Privacy Preserving Technologies \& Security in AI (NTU), \tabref{privacy_ai_security}, Advanced Real-World Networks (CMU), \tabref{advanced_networks}, Networked Systems Course (CMU), \tabref{networked_systems_security}, Information Security Policy and Management (CMU), \tabref{security_policy_management}.

For each course analyzed, we derived topics from the course descriptions using PreprocessLM, labeled them using ClassifyLM and present the final result of this process under the column \CLLM{} in the table. Then, Control Group A, consisting of 34 cybersecurity experts, and a group of three cybersecurity experts previously familiar with the CSEC2017 framework (noted as `Expert Group X') were consulted to independently label the topics. The numbers listed in the tables down below represent these annotations, where the corresponding KAs for each number can be found in \tabref{KAs}. 

\begin{table}[ht!]
\centering
\caption{Topics extracted from the Ethical Hacking Course (KTH) annotated by 3 annotators from Control Group A, all annotators in Expert Group X, and \CLLM{}}
\label{tab:ethical_hacking}
\begin{tabular}{lccccccc}
\toprule
  & \multicolumn{3}{c}{\textbf{Control Group A}} & \multicolumn{3}{c}{\textbf{Expert Group X}} & \textbf{\CLLM{}} \\
\cmidrule(l{1mm}r{1mm}){2-4}
\cmidrule(l{1mm}r{1mm}){5-7}
\textbf{Topics} & \textbf{\#1} & \textbf{\#2} & \textbf{\#3} & \textbf{$X_1$} & \textbf{$X_2$} & \textbf{$X_3$} &  \\ \midrule
Ethical Hacking & 1,6 & 6,8 & 2,5,6 & 0--8 & 2,7,8 & 0,2--6 & 8 \\ 
\graybg Network and vulnerability scanning & 4 & 3,4,5,7 & 1,4 & 4,5 & 5 & 0,4 & 4 \\ 
Exploit development platforms & 2,7 & 3 & 2,3 & 2 & 2 & 2,3,4 & 2 \\ 
\graybg Command and control & 0 & 2 & 2,3,6 & 0 & 7 & 0,2,4,5 & 0 \\ 
Password cracking & 1 & 5,6 & 1,2,3 & 0,1 & 5 & 1,2,6 & 6 \\ 
\graybg Independent attack project & 1,3 & -- & 7 & 0 & 5 & 0 & 5 \\ 
Virtual environment setup & 4 & 7 & 1,2 & 4,5 & 0 & 0 & 4 \\ \bottomrule
\end{tabular}
\end{table}

\begin{table}[ht!]
\centering
\caption{Topics extracted from the Language-Based Security Course (KTH) annotated by 3 annotators from Control Group A, all annotators in Expert Group X, and \CLLM{}}
\label{tab:language_based_security}
\resizebox{.98\textwidth}{!}{
\begin{tabular}{m{6.5cm}ccccccc}
\toprule
  & \multicolumn{3}{c}{\textbf{Control Group A}} & \multicolumn{3}{c}{\textbf{Expert Group X}} & \textbf{\CLLM{}} \\
\cmidrule(l{1mm}r{1mm}){2-4}
\cmidrule(l{1mm}r{1mm}){5-7}
\textbf{Topics} & \textbf{\#1} & \textbf{\#2} & \textbf{\#3} & \textbf{$X_1$} & \textbf{$X_2$} & \textbf{$X_3$} &  \\ \midrule
Language-Based Security & 1,6 & 2 & -- & 0,2 & 6 &  0,1,2,6,7 & 5 \\
\graybg Fundamental principles, models and concepts for computer security & 0 & 2 & 1,4,5,7,8 & 0--5 & 1,3 & 0--8 & 5 \\ 
Software security by information flow control & 3 & 4 & 1,2,4 & 2 & 2 & 2 & 2 \\ 
\graybg Web application and database security & 2 & 2 & 1,4,5,6,7 & 0,2 & 2,5 & 1,2,4,5,7 & 7 \\ 
Security for mobile applications & 5,6 & 2 & 1,3,4,5,6 & 4 & 2,4 & 2,5,6 & 7 \\ 
\graybg Hot topics in computer security & 1 & -- & 0,5,8 & 0--8 & 1,3,4,5 & 0--8 & 7 \\ 
State-of-the-art in programming language for security & 1,6 & 2 & 2,3,4 & 2 & 2 & 2 & 7 \\ \bottomrule
\end{tabular}
}
\end{table}

\begin{table}[ht!]
\centering
\caption{Topics extracted from the Network Security Course (NTU) annotated by 3 annotators from Control Group A, all annotators in Expert Group X, and \CLLM{}}
\label{tab:network_security}
\resizebox{.98\textwidth}{!}{
\begin{tabular}{lccccccc}
\toprule
  & \multicolumn{3}{c}{\textbf{Control Group A}} & \multicolumn{3}{c}{\textbf{Expert Group X}} & \textbf{\CLLM{}} \\
\cmidrule(l{1mm}r{1mm}){2-4}
\cmidrule(l{1mm}r{1mm}){5-7}
\textbf{Topics} & \textbf{\#1} & \textbf{\#2} & \textbf{\#3} & \textbf{$X_1$} & \textbf{$X_2$} & \textbf{$X_3$} &  \\ \midrule
Network Security & 5 & 3,4,5,7 & 1,2,5,6,7  & 4 & 4,5 &  4 & 7 \\ 
\graybg Introduction to Networking & 5  & 1,2,3,4,5 & 4  & 0 & 0 & 4 & 4 \\ 
Security in Networks & 5  & 3,4,5,7 & 1,2,5,6  & 4 & 4,5 & 4 & 7 \\
\graybg Network Programming & 5 & 4,5 & 2 & 0,4 & 2,4 & 2,4 & 7 \\ 
Packet Trace Analysis with Wireshark & 5 & 4 & 1,4,7 & 4 & 4 & 3,4 & 4 \\
\graybg TLS and Security in Networked Systems & 5 & 1,2,4,5 & 1,4,5 & 4 & 4,5 & 4 & 4 \\ \bottomrule

\end{tabular}
}
\end{table}

\begin{table}[ht!]
\centering
\caption{Topics extracted from the Software Security Course (NTU) annotated by 3 annotators from Control Group A, all annotators in Expert Group X, and \CLLM{}}
\label{tab:software_security}
\resizebox{.98\textwidth}{!}{
\begin{tabular}{lccccccc}
\toprule
  & \multicolumn{3}{c}{\textbf{Control Group A}} & \multicolumn{3}{c}{\textbf{Expert Group X}} & \textbf{\CLLM{}} \\
\cmidrule(l{1mm}r{1mm}){2-4}
\cmidrule(l{1mm}r{1mm}){5-7}
\textbf{Topics} & \textbf{\#1} & \textbf{\#2} & \textbf{\#3} & \textbf{$X_1$} & \textbf{$X_2$} & \textbf{$X_3$} &  \\ \midrule
Software Security & 3,4,5,7 & -- & 2 & 2 & 2 & 2 & 2 \\ 
\graybg Challenges in software security & 1--5 & 8 & 2 & 2 & 2 & 2 & 2 \\ 
Principles of secure software development & 3--5,7 & 2 & 0,2,5 & 2 & 2 & 2 & 2 \\ 
\graybg Mechanisms for secure coding & 4 & 2,5 & 5 & 2 & 2 & 2 & 2 \\ 
Tools for software security & 2,4 & 1 & 2 & 2 & 2 & 2 & 2 \\ 
\graybg Causes of vulnerabilities & 1,2,4,5 & 3,6 & 0,6,7 & 0--6 & 2--5 & 0--8 & 7 \\ 
Strategies to avoid vulnerabilities & 1--3 & 2 & 3,5,7 & 0--6 & 2,3 & 2,5,6,7,8 & 7 \\ 
\graybg Defenses against attacks & 1--5,7 & 1,6 & 3,5,7 & 0--6 & 5,7 & 0--8 & 5\\ 
Secure programming practices & 2 & 3 & 2,3,5,6,8 & 2 & 2 & 2 & 2 \\ 
\graybg Language-specific secure coding & 2,4 & 3 & 2,3,5,7 & 2 & 2 & 2 & 2 \\ 
System-level defenses & 1,2,4,5 & 5 & 1--5 & 5 & 5 & 5 & 5 \\ 
\graybg Architectural approaches for security & 1,3--5,7,8 & 2,4 & 0,3--8 & 4 & 5 & 0,5,6,7,8 & 2 \\ 
Run-time enforcement techniques & 2--6 & 1,5 & 0,4,6--8 & 4 & 2,3 & 2,3,5 & 7 \\ 
\graybg Software analysis tools & 1,2,4,6,7 & 2 & 0--2,4--8 & 2 & 0,2 & 0,2,3 & 2 \\ 
Vulnerability detection methods & 1--3,5--8 & 3 & 0,2,5--8 & 3 & 5 & 0,2--6 & 2,3 \\ 
\graybg Application of security tools in scenarios & 0,2,5--8 & -- & 0--2,5--8 & 3 & 5 & 0 & 7 \\ \bottomrule
\end{tabular}
}
\end{table}

\begin{table}[ht!]
\centering
\caption{Topics extracted from Privacy Preserving Technologies \& Security in AI Course (NTU) annotated by 3 annotators from Control Group A, all annotators in Expert Group X, and \CLLM{}}
\label{tab:privacy_ai_security}
\resizebox{.98\textwidth}{!}{
\begin{tabular}{lccccccc}
\toprule
  & \multicolumn{3}{c}{\textbf{Control Group A}} & \multicolumn{3}{c}{\textbf{Expert Group X}} & \textbf{\CLLM{}} \\
\cmidrule(l{1mm}r{1mm}){2-4}
\cmidrule(l{1mm}r{1mm}){5-7}
\textbf{Topics} & \textbf{\#1} & \textbf{\#2} & \textbf{\#3} & \textbf{$X_1$} & \textbf{$X_2$} & \textbf{$X_3$} &  \\ \midrule
Privacy Preserving Technologies \& Security in AI & 1,6 & 0--3,5,7,8 & 0--2,6--8 & 0,2,3,6 & 5,6,8 & 0,1,6 & 1,6--8 \\ 
\graybg Fully Homomorphic Encryption & 1 & 0,1,4 & 1,3,5 & 1 & 1 & 1 & 1 \\ 
Secure Multi-computation & 1,2,4 & 0,2,3,5,7 & 1,3,5,7 & 1 & 1 & 1 & 1 \\ 
\graybg Searchable Encryption & 1 & 0,1,4 & 1,2,5 & 1 & 1 & 1 & 1 \\ 
Data Sharing in Machine Learning & 1 & 0,4,7,8 & 0--2,4,5,7 & 0,1 & 0 & 1,6 & 4 \\ \bottomrule
\end{tabular}
}
\end{table}

\begin{table}[ht!]
\centering
\caption{Topics extracted from Advanced Real-World Networks Course (CMU) annotated by 3 annotators from Control Group A, all annotators in Expert Group X, and \CLLM{}}
\label{tab:advanced_networks}
\begin{tabular}{lccccccc}
\toprule
  & \multicolumn{3}{c}{\textbf{Control Group A}} & \multicolumn{3}{c}{\textbf{Expert Group X}} & \textbf{\CLLM{}} \\
\cmidrule(l{1mm}r{1mm}){2-4}
\cmidrule(l{1mm}r{1mm}){5-7}
\textbf{Topics} & \textbf{\#1} & \textbf{\#2} & \textbf{\#3} & \textbf{$X_1$} & \textbf{$X_2$} & \textbf{$X_3$} &  \\ \midrule
Advanced Real-World Networks & 4,5 & 1--8 & 1,4,5 & 0,4 & 0 & 4 & 4 \\ 
\graybg 4G and 5G network infrastructures & 4 & 1--8 & 0,3--5 & 0,4 & 0 & 1,4,5 & 4 \\ 
IPv6 & 2,5 & 0 & 0,1,4 & 4 & 0 & 4 & 4 \\ 
\graybg SDN and VFN & 2 & 1,2,4,7 & 0,1,2,3,5 & 4 & 0 & 2,4 & 4 \\ 
Data centers & 1 & 1,3--6 & 0--7 & 0 & 0 & 1--5 & 1 \\ 
\graybg Mesh networks & 3,4 & 1--5 & 0,4 & 4 & 0 & 4 & 4 \\ 
Embedded networks & 3,4 & 1--8 & 0,4 & 4 & 0 & 2,4 & 4 \\ \bottomrule
\end{tabular}
\end{table}

\begin{table}[ht!]
\centering
\caption{Topics extracted from Security in Networked Systems Course (CMU) annotated by 3 annotators from Control Group A, all annotators in Expert Group X, and \CLLM{}}
\label{tab:networked_systems_security}
\resizebox{.98\textwidth}{!}{
\begin{tabular}{m{6.9cm}ccccccc}
\toprule
  & \multicolumn{3}{c}{\textbf{Control Group A}} & \multicolumn{3}{c}{\textbf{Expert Group X}} & \textbf{\CLLM{}} \\
\cmidrule(l{1mm}r{1mm}){2-4}
\cmidrule(l{1mm}r{1mm}){5-7}
\textbf{Topics} & \textbf{\#1} & \textbf{\#2} & \textbf{\#3} & \textbf{$X_1$} & \textbf{$X_2$} & \textbf{$X_3$} &  \\ \midrule
Security in Networked Systems & 1 & 1--5 & 1 & 4 & 4,5 & 4 & 7 \\ 
\graybg Network and transport-layer attacks and defenses & 4 & 1--5 & 4 & 4 & 4,5 & 4 & 4 \\ 
Network intrusion detection & 5,6,7 & 1--5 & 1,4 & 4 & 4,5 & 4 & 4 \\ 
\graybg Denial of service (DoS) and distributed denial-of-service (DDoS) detection and reaction & 4,5,8 & 4 & 1,4 & 4 & 4,5 & 4 & 4 \\ 
Worm and virus propagation & 2,3 & 6 & 2,3,5,7 & 4 & 2,4,5 & 0,2--6 & 1 \\ 
\graybg Tracing the source of attacks & 0-8 & 0--2,4-6-,8 & 1,2 & 0-6 & 5 & 0,2,4,5 & 1 \\ 
Traffic analysis & 4 & 0,1,4,8 & 4 & 4 & 1,4 &0,4  & 7 \\ 
\graybg Techniques for hiding the source or destination of network traffic & 1,2,5 & 0,1,4,6,8 & 2,4,5 & 4 & 4 & 4 & 4 \\ 
Secure routing protocols & 4 & 0,1,2,4,8 & 4 & 4 & 4 & 4 & 4 \\ 
\graybg Content poisoning attacks & 1,2,3 & 1--4 & 1 & 4 & 2--4 & 2,4 & 1 \\ 
Advanced techniques for reacting to network attacks & 4,8 & 0--6 & 0,6--8 & 4 & 4,5 & 0,4 & 4 \\ \bottomrule
\end{tabular}
}
\end{table}

\begin{table}[ht!]
\centering
\caption{Topics extracted from Information Security Policy and Management Course (CMU) annotated by 3 annotators from Control Group A, all annotators in Expert Group X, and \CLLM{}}
\label{tab:security_policy_management}
\resizebox{.98\textwidth}{!}{
\begin{tabular}{m{7.4cm}ccccccc}
\toprule
  & \multicolumn{3}{c}{\textbf{Control Group A}} & \multicolumn{3}{c}{\textbf{Expert Group X}} & \textbf{\CLLM{}} \\
\cmidrule(l{1mm}r{1mm}){2-4}
\cmidrule(l{1mm}r{1mm}){5-7}
\textbf{Topics} & \textbf{\#1} & \textbf{\#2} & \textbf{\#3} & \textbf{$X_1$} & \textbf{$X_2$} & \textbf{$X_3$} &  \\ \midrule
Information Security Policy and Management & 0,1,5,6,8 & 7,8 & 7 & 7 & 5,7 & 0,7,8 & 7 \\ 
\graybg Security marketplace overview & 0,2,8 & 3 & 0 & 7 & 5,7 & 0 & 7 \\ 
Decision making with multiple parties involved & 5,7,8 & 6,8 & 8 & 7 & 0 & 0,1,7,8 & 7,8 \\ 
\graybg Role of policy in information security & 1,7 & 8 & 7 & 7,8 & 5,7,8 & 7 & 7,8 \\ 
Intra-organization policies & 4,7 & 4,8 & 3,5,6 & 7 & 0 & 7 & 7 \\ 
\graybg Market and competition impact on security provision & 0,7,8 & 1,2 & 8 & 7 & 5,7,8 & 0,7,8 & 7 \\ 
Causes of market failure (externalities) & 0,5,7,8 & 0,7 & 6--8 & 0,7 & 0 & 0,7,8 & 7\\ 
\graybg Policy tools to mitigate market failure & 0,5 & 0,8 & 3,7,8 & 7 & 0 & 0,7 & 7,8 \\ 
Key laws and regulations on product liability and security standards & 5 & 3,7,8 & 1,2 & 7,8 & 8 &  0,3,4,5,7,8& 7 \\ 
\graybg Overview of security industry trends, technologies, vendor and user strategies & 0 & 0--5 & 0,4 & 7 & 8 & 0 & 7 \\ 
Managerial and policy issues in information security provision & 5,7 & 0,1,3,6 & 0,2 & 7 & 7,8 & 7 & 7 \\ 
\graybg When and how policy intervention is needed &-- & 0,7,8 & 6,7 & 7,8 & 8 & 0,7,8 & 7,8 \\ \bottomrule
\end{tabular}
}
\end{table}

\section{Zero-shot classification prompt}
\label{app:zero_shot_classification}

We employed zero-shot prompting of ChatGPT-mini-4o, and DeepSeek-V3 as a baseline for comparison with the proposed ClassifyLM structure (see \secref{baselines}). Zero-shot prompting requires the selected model to perform Knowledge Area annotation based solely on the instructions and definitions provided in the prompt, without any prior examples. 
The prompt provided to both ChatGPT-mini-4o and DeepSeek-V3 is detailed in \figref{llmprompt2}. 


\begin{figure}[ht!]
\centering
\begin{tcolorbox}[colback=gray!5!white,colframe=black!75!black,
  title=Prompt Structure for Knowledge Area Classification]
\textbf{System Prompt:}\\
\texttt{
"You are a helpful AI assistant.\\
Instructions:\\
a. Carefully read the knowledge statement.\\
b. Choose one or more of the following (0, 1, 2, 3, 4, 5, 6, 7, 8).\\
c. Do NOT include any explanation or additional text in the response."}

\vspace{0.5em}
\textbf{User Message:}\\
\texttt{
"\#Question: Classify the following statement \{knowledge\} into one or multiple of the following knowledge areas:\\
Options: \{"0": "miscellaneous (this includes Computer Science, Business and Law, Communication and Networking, Information Technology, Cyberspace Practice, Pedagogy, and Intelligence)",
           "1": "data security", 
           "2": "software security",
           "3": "component security", 
           "4": "connection security", 
           "5": "system security", 
           "6": "human security", 
           "7": "organizational security",
           "8": "societal security"\}"}
\end{tcolorbox}
\caption{The prompt template used for annotating knowledge statements with Knowledge Areas using Zero-Shot prompting of DeepSeek-V3 or ChatGPT-4o-mini. The placeholder \{knowledge\} is replaced by a NICE Knowledge Description or a synthetically derived CSEC2017 topic.}
\label{fig:llmprompt2}
\end{figure}

\section{KA distribution NICE Job Roles}
\label{app:KA_job_roles}

The relative importance of each of the nine Knowledge Areas for all work role categories defined in the NICE framework v2.0.0 is shown in \figref{job_role_KAs_combined} and discussed in \secref{job_to_KA}. The `relevance' of KAs was derived by collecting the KDs associated with a given job role, annotating them with corresponding KAs using ClassifyLM, and computing the resulting  distribution of KAs. 
For completeness, this appendix provides the relative importance of the KAs for \textit{all} individual NICE work roles. These detailed results are displayed in \tabref{nice_work_roles}. 

\begin{table}[ht!]
\centering
\caption{Relative importance of CSEC2017 Knowledge Areas for all Work Roles and Work Role Categories contained in the NICE Framework v2.0.0. Knowledge Descriptions contained in each work role were annotated using \CLLM{} and the numbers represent percentage-wise contributions, rounded to the nearest integer.}
\label{tab:nice_work_roles}
\resizebox{\textwidth}{!}{%
\begin{tabular}{lcccccccccc}
\toprule
\textbf{Work Role } & \textbf{KA0} & \textbf{KA1} & \textbf{KA2} & \textbf{KA3} & \textbf{KA4} & \textbf{KA5} & \textbf{KA6} & \textbf{KA7} & \textbf{KA8} \\
\midrule
OVERSIGHT and GOVERNANCE (OG) & 16 & 7 & 3 & 1 & 9 & 3 & 7 & 41 & 12 \\
\graybg \quad Communications Security (COMSEC) Management & 10 & 12 & 5 & 5 & 5 & 4 & 8 & 38 & 11 \\
\quad Cybersecurity Policy and Planning & 15 & 6 & 2 & 0 & 8 & 2 & 6 & 44 & 17 \\
\graybg \quad Cybersecurity Workforce Management & 19 & 4 & 3 & 0 & 7 & 0 & 4 & 51 & 12 \\
\quad Cybersecurity Curriculum Development & 33 & 6 & 0 & 0 & 11 & 2 & 8 & 29 & 12 \\
\graybg \quad Cybersecurity Instruction & 26 & 8 & 0 & 0 & 17 & 3 & 8 & 26 & 10 \\
\quad Cybersecurity Legal Advice & 17 & 12 & 2 & 0 & 5 & 0 & 12 & 36 & 17 \\
\graybg \quad Executive Cybersecurity Leadership & 8 & 5 & 2 & 0 & 15 & 3 & 5 & 46 & 15 \\
\quad Privacy Compliance & 10 & 11 & 1 & 1 & 11 & 4 & 13 & 31 & 18 \\
\graybg \quad Product Support Management & 18 & 4 & 4 & 3 & 7 & 3 & 4 & 48 & 10 \\
\quad Program Management & 16 & 5 & 5 & 3 & 8 & 3 & 5 & 45 & 11 \\
\graybg \quad Secure Project Management & 18 & 4 & 4 & 3 & 7 & 3 & 4 & 45 & 10 \\
\quad Security Control Assessment & 23 & 9 & 6 & 3 & 11 & 5 & 6 & 31 & 7 \\
\graybg \quad Systems Authorization & 15 & 13 & 3 & 0 & 8 & 5 & 6 & 39 & 11 \\
\quad Systems Security Management & 14 & 7 & 6 & 4 & 12 & 6 & 5 & 39 & 7 \\
\graybg \quad Technology Portfolio Management & 7 & 7 & 2 & 0 & 9 & 0 & 7 & 51 & 16 \\
\quad Technology Program Auditing & 16 & 6 & 2 & 0 & 6 & 0 & 6 & 51 & 14 \\
\midrule
DESIGN and DEVELOPMENT (DD) & 14 & 6 & 10 & 4 & 11 & 9 & 5 & 32 & 8 \\
\graybg \quad Cybersecurity Architecture & 17 & 7 & 9 & 4 & 16 & 13 & 5 & 25 & 5 \\
\quad Enterprise Architecture & 14 & 6 & 7 & 5 & 14 & 13 & 6 & 28 & 6 \\
\graybg \quad Secure Software Development & 13 & 3 & 19 & 4 & 11 & 11 & 4 & 28 & 9 \\
\quad Secure Systems Development & 17 & 8 & 12 & 3 & 13 & 8 & 6 & 26 & 7 \\
\graybg \quad Software Security Assessment & 12 & 3 & 19 & 5 & 10 & 9 & 4 & 29 & 9 \\
\quad Systems Requirements Planning & 14 & 9 & 6 & 4 & 8 & 8 & 8 & 34 & 9 \\
\graybg \quad Systems Testing and Evaluation & 14 & 6 & 9 & 5 & 8 & 9 & 6 & 34 & 10 \\
\quad Technology Research and Development & 18 & 8 & 8 & 5 & 16 & 7 & 3 & 25 & 10 \\
\graybg \quad Operational Technology (OT) Cybersecurity Engineering & 5 & 7 & 0 & 0 & 7 & 5 & 5 & 62 & 9 \\
\midrule
IMPLEMENTATION and OPERATION (IO) & 20 & 11 & 4 & 2 & 12 & 5 & 8 & 29 & 10 \\
\graybg \quad Data Analysis & 20 & 13 & 7 & 1 & 7 & 8 & 8 & 25 & 11 \\
\quad Database Administration & 22 & 17 & 0 & 0 & 7 & 1 & 12 & 29 & 12 \\
\graybg \quad Knowledge Management & 31 & 8 & 2 & 0 & 8 & 0 & 8 & 31 & 13 \\
\quad Network Operations & 20 & 10 & 4 & 4 & 21 & 5 & 7 & 23 & 8 \\
\graybg \quad Systems Administration & 12 & 8 & 5 & 4 & 21 & 7 & 6 & 28 & 8 \\
\quad Systems Security Analysis & 18 & 9 & 6 & 4 & 14 & 8 & 6 & 28 & 7 \\
\graybg \quad Technical Support & 18 & 10 & 1 & 0 & 9 & 4 & 10 & 37 & 10 \\
\midrule
PROTECTION and DEFENSE (PD) & 15 & 11 & 3 & 1 & 16 & 5 & 6 & 33 & 9 \\
\graybg \quad Defensive Cybersecurity & 17 & 9 & 6 & 3 & 20 & 8 & 5 & 26 & 6 \\
\quad Digital Forensics & 14 & 23 & 5 & 4 & 11 & 9 & 3 & 22 & 10 \\
\graybg \quad Incident Response & 10 & 10 & 1 & 0 & 19 & 3 & 10 & 37 & 10 \\
\quad Infrastructure Support & 11 & 11 & 0 & 0 & 22 & 3 & 9 & 34 & 11 \\
\graybg \quad Insider Threat Analysis & 15 & 10 & 0 & 0 & 8 & 3 & 5 & 51 & 8 \\
\quad Threat Analysis & 32 & 5 & 2 & 0 & 18 & 3 & 5 & 26 & 8 \\
\graybg \quad Vulnerability Analysis & 9 & 10 & 4 & 0 & 14 & 6 & 9 & 36 & 14 \\
\midrule
INVESTIGATION (IN) & 12 & 21 & 3 & 2 & 9 & 7 & 5 & 28 & 12 \\
\graybg \quad Cybercrime Investigation & 12 & 18 & 1 & 0 & 9 & 4 & 6 & 35 & 15 \\
\quad Digital Evidence Analysis & 13 & 25 & 5 & 4 & 10 & 10 & 4 & 21 & 10 \\
\bottomrule
\end{tabular}%
}
\end{table}

\section{Knowledge Area 0: Miscellaneous}\label{KnowledgeArea0:Miscellaneous}

In \citep{ramezanian2024cybersecurity} the authors identified seven knowledge units in the knowledge area 0: Miscellaneous. The description of these knowledge units are as follows.

\begin{itemize}
    \item \textit{Computer Science:} includes topics related to Programming and Software Development, System Architecture, Database Management, Mathematics, Information Systems, and Human and Organizational Aspects of Information Technologies. 
    \item \textit{Business and Law:} includes topics related to Regulatory Frameworks, Organizational Operations \& Policies, Stakeholder Relationships, and Management \& Human Resources.
    \item \textit{Communication and Networking:} includes topics related to Telecommunications Fundamentals, Network Architectures Fundamentals, Network Management, Communication Systems, and Information Gathering and Dissemination.
    \item \textit{Information Technology:} includes topics related to Technology Infrastructure, Content Systems, Technology Trends \& Integration, and Media, Communication, \& Dissemination. 
    \item \textit{Cyberspace Practice:} includes topics related to Adversarial Models, Cyber Attacks \& Vulnerabilities, Cyber Operations \& Practices, Core Cyber Concepts \& Principles, and Security Tools, Vendors, \& Defensive Technologies.
    \item \textit{Pedagogy:} includes topics related to Learning Frameworks, Educational Technology, Training Policies, Organizational Education Practices, Assessment Techniques, Experiential \& Applied Learning, and Psychology, Behavior, \& Human Factors.
    \item \textit{Intelligence:} includes topics related to Intelligence Fundamentals, Intelligence Frameworks, Intelligence Collection, Intelligence Management, Targeting \& Operational Intelligence, Intelligence Production, Intelligence Dissemination, Intelligence Planning, Resource Prioritization, Partner \& Organizational Intelligence Integration, and Covert \& Specialized Intelligence Techniques. 
\end{itemize}

\end{document}